\documentclass[manuscript]{acmart}

\usepackage{enumitem}
\usepackage[justification=centering]{caption}
\usepackage{tcolorbox} % Add this to your preamble if not already
\usepackage{multirow}
\usepackage{makecell}
\usepackage{url}
\usepackage{hyperref}

\usepackage{listings}
\usepackage{xcolor}

\tcbuselibrary{skins}
\newtcolorbox{rqbox}[2]{
  enhanced,
  colback=black!7,
  colframe=black!7,
  boxrule=0pt,
  arc=2pt,
  left=10pt, right=10pt, top=8pt, bottom=8pt,
  attach boxed title to top left={yshift=-2mm, xshift=8pt},
  boxed title style={
    colback=black!55,
    colframe=black!55,
    sharp corners,
    arc=2pt,
    fontupper=\sffamily\normalsize\bfseries\color{white} % badge stays small
  },
  title={#1\quad{\normalfont\sffamily\small\color{white}#2}},  % title larger
}

\newtcolorbox{insightbox}{
  enhanced,
  colback=black!7,
  colframe=black!7,
  boxrule=0pt,
  borderline west={3pt}{0pt}{black!70},
  arc=2pt,
  left=10pt, right=10pt, top=6pt, bottom=6pt,
  before upper={%
    {\sffamily\bfseries\small\MakeUppercase{Summary}}%
    \par\vspace{2pt}%
    {\color{black!60}\hrule height 0.8pt}\par\vspace{4pt}%
  }
}

\AtBeginDocument{%
  }

\setcopyright{acmlicensed}
\copyrightyear{2026}
\acmYear{2026}
\acmDOI{XXXXXXX.XXXXXXX}

\usepackage{algorithm}
\usepackage{algorithmic}

\usepackage{graphicx}
\usepackage[table,xcdraw]{xcolor}

\usepackage{url}

\usepackage{changepage}
\usepackage{framed}
\makeatletter
{\par\unskip\endMakeFramed}
\makeatother
\definecolor{formalshade}{HTML}{F9E3DF}

\definecolor{darkblue}{rgb}{0.2, 0.2, 0.2}

\newtcolorbox{summarybox}[1][]{
  colback=blue!5!white,    % background color
  colframe=blue!75!black,  % border color
  coltitle=white,          % title text color
  fonttitle=\bfseries,     % title font style
  title=Summary,           % default title
  sharp corners,
  boxrule=0.8pt,
  left=6pt,right=6pt,top=6pt,bottom=6pt,
  #1
}

\usepackage{pgfplots}
\usepackage{pgfplotstable}
\pgfplotsset{compat=1.18}
\usepackage{xcolor}
\usepackage{subcaption}
\usepackage{float}

\definecolor{colAutoWin} {RGB}{55,138,221}   % blue  – auto wins
\definecolor{colManWin}  {RGB}{180,178,169}  % gray  – manual wins
\definecolor{colTie}     {RGB}{136,135,128}  % dark gray – tie
\definecolor{colDiag}    {RGB}{226,75,74}    % red   – diagonal

\usepackage{pgfplots}
\usepackage{pgfplotstable}
\pgfplotsset{compat=1.18}
\usepackage{booktabs}
\usepackage{xcolor}
\usepackage{caption}
\usepackage{subcaption}
\usepackage{float}
 
\definecolor{colManual}{RGB}{180,178,169}   % gray  – manual (0)
\definecolor{colAuto}  {RGB}{55,138,221}    % blue  – automated (1)
\definecolor{colAutoD} {RGB}{12,68,124}     % dark blue (VFC auto)
\definecolor{colManualD}{RGB}{136,135,128}  % dark gray (VFC manual)
\definecolor{colSig}   {RGB}{226,75,74}     % red diagonal / significance
 
\pgfplotsset{
  rq1 base/.style={
    width=0.92\linewidth,
    height=6cm,
    font=\small,
    grid=major,
    grid style={dotted, gray!40},
    tick align=inside,
    tick style={color=black},
    axis line style={black},
    legend style={
      font=\footnotesize,
      draw=gray!50,
      fill=white,
      at={(0.97,0.97)},
      anchor=north east,
    },
    ylabel style={align=center},
  }
}

\definecolor{colDiag}{RGB}{226,75,74}     % red diagonal
 
\definecolor{budA}{RGB}{210,230,245}      % 5%  — very light blue
\definecolor{budB}{RGB}{130,190,225}      % 7.5% — light-mid blue
\definecolor{budC}{RGB}{55,140,200}       % 10%  — mid blue
\definecolor{budD}{RGB}{20,80,155}        % 15%  — dark blue
\definecolor{budE}{RGB}{8,30,90}          % 20%  — very dark navy
 
\definecolor{negGray} {RGB}{241,239,232}
\definecolor{smBlue}  {RGB}{230,241,251}
\definecolor{medGreen}{RGB}{234,243,222}
\definecolor{lgAmber} {RGB}{250,238,218}
\definecolor{vlgRed}  {RGB}{252,235,235}
 
\usepackage{booktabs}
\usepackage{multirow}
\usepackage{array}
\usepackage{geometry}
\usepackage{caption}
\usepackage{microtype}
\usepackage{threeparttable}

\begin{document}

% \title{On the Impact of Data Labeling Strategies for Just-In-Time Vulnerability Prediction}
\title{One Strategy Does Not Fit All: Rethinking Data Labeling for Just-in-Time Vulnerability Prediction}

\author{Quan Tran}
\affiliation{
  \institution{Phenikaa University}
  \country{Hanoi, Vietnam}
}
\email{minhquantsy@gmail.com}
\orcid{0009-0000-9489-6714}

\author{Duong Nguyen}
\affiliation{
  \institution{Singapore Management University}
  \country{Singapore}
}
\email{dnguyen@smu.edu.sg}
\orcid{0009-0001-4715-3106}

\author{Thanh Le-Cong}
\affiliation{
  \institution{Singapore University of Technology and Design}
  \country{Singapore}
}
\email{congthanh_le@sutd.edu.sg}
\orcid{0000-0002-9566-324X}

\author{Triet Huynh Minh Le}
\affiliation{
  \institution{Adelaide University}
  \country{South Australia, Australia}
}
\email{triet.h.le@adelaide.edu.au}
\orcid{0000-0003-1935-037X}

\author{M. Ali Babar}
\affiliation{
  \institution{Adelaide University}
  \country{South Australia, Australia}
}
\email{ali.babar@adelaide.edu.au}
\orcid{0000-0001-9696-3626}

\author{Hoang Dieu Vu}
\affiliation{
  \institution{Phenikaa University}
  \country{Hanoi, Vietnam}
}
\email{dieu.vuhoang@phenikaa-uni.edu.vn}
\orcid{0000-0002-8732-8302}

\author{Quyet-Thang Huynh}
\authornote{Corresponding author.}
\affiliation{
  \institution{Hanoi University of Science and Technology}
  \country{Hanoi, Vietnam}
}
\email{thang.huynhquyet@hust.edu.vn}
\orcid{0000-0002-0788-6380}

\renewcommand{\shortauthors}{Tran et al.}

\begin{abstract}
Just-In-Time Vulnerability Prediction (JIT-VP) identifies security risks in code changes at commit time, enabling early feedback before vulnerabilities enter software systems. However, its practical reliability is challenged by the difficulty of constructing accurate labels. Multiple strategies with different trade-offs have been proposed to identify vulnerability-fixing commits (VFCs) and vulnerability-introducing commits (VICs) for dataset construction, leading to substantially different datasets. This makes it difficult to separate model performance from dataset construction. Moreover, real-world quality assurance requires not only classification but also prioritizing high-risk commits and localizing vulnerable causes. This paper presents an extensive empirical study of how labeling strategies affect these downstream tasks. We constructed 28 datasets using four VFC and four VIC labeling strategies across OpenSSL and Linux kernel, then evaluated six representative JIT-VP models for classification, prioritization, and localization. Our results show that labeling effectiveness is strongly task-dependent. Manual VFC labeling combined with vulnerability-specific VIC labeling provides the most reliable performance for classification and prioritization, significantly outperforming 11 of 15 alternatives. Meanwhile, precision-first VFC labeling combined with LLM-assisted VIC labeling performs best for localization, significantly outperforming 7 of 15 alternatives. These findings demonstrate that no single labeling strategy is universally optimal and highlight the importance of task-oriented dataset construction.

\end{abstract}
% CCS Concepts


\maketitle

\section{Introduction}
\label{sec:introduction}

The rapid adoption of AI-powered coding agents has substantially accelerated software development, leading to a substantial growth in code commits. 
For example, GitHub recorded only one billion commits in 2025, yet by early April 2026, it was processing 275 million commits per week, a rate that would amount to approximately 14 billion commits over the year if sustained~\cite{daigle2026}. 
Such a massive volume of code changes across different software versions and deployment environments increases the likelihood of failures caused by inconsistencies between software components and patches, as well as conflicts among the patches themselves.
Consequently, human reviewers are confronted with an escalating challenge in identifying risky code changes that could compromise system stability and security. 

Just-In-Time Vulnerability Prediction (JIT-VP)~\cite{perl2015vccfinder} has emerged as a promising approach for assisting human reviewers and improving software quality assurance practices.  
By learning from historical vulnerability data, JIT-VP techniques can learn patterns associated with risky code modifications, thereby providing developers with early feedback.  
As a result, integrating JIT-VP into the software development lifecycle can improve the efficiency of security review processes while enhancing the resilience of software systems against emerging threats. However, their practical readiness remains a concern due to the complexity of real-world software development~\cite{nguyen2025toward}.

One key challenge in JIT-VP is constructing reliable labels for vulnerability-introducing commits (VICs). This typically requires two steps. First, researchers must identify the vulnerability-fixing commit (VFC), but vulnerability records often lack a direct link to the patch, and commit metadata may not reveal that it is a security fix~\cite{nguyen2025mapping}. Second, they must determine which earlier commit introduced the vulnerability. A VFC shows what was changed to fix the vulnerability, but it does not establish when the vulnerable behavior was introduced. For example, tracing a fixed line through the commit history may identify a commit that merely modified already-vulnerable code~\cite{bao2022v,sliwerski2005changes}. Thus, even a correctly identified VFC can lead to an incorrect VIC label.

Consequently, various strategies have been proposed to identify VFCs and VICs to construct JIT-VP labels, as discussed in Section~\ref{sec:design}. 
These strategies entail different trade-offs among labeling accuracy, coverage, scalability, cost, and semantic detail. 
The choice of labeling strategy can therefore substantially influence the resulting training data, as illustrated in 
Section~\ref{sec:motivation}, and consequently affect the measured performance of JIT-VP models. Despite the importance of this design choice, existing studies (e.g.,~\cite{perl2015vccfinder, nguyen2024code, nguyen2025toward}) often employ different data construction pipelines, making their results difficult to interpret independently of the underlying label construction process. Consequently, the impact of different labeling strategies on downstream JIT-VP performance remains under-explored.

Moreover, practical software development increasingly demands more from JIT-VP models than binary vulnerability prediction. The scale and complexity of modern software systems, combined with rapid development cycles, limit the amount of effort developers can devote to inspecting model predictions. Therefore, simply determining whether a commit is vulnerable may be insufficient in practice. JIT-VP approaches should also help developers prioritize potentially vulnerable commits under limited inspection effort and provide fine-grained information about the code changes associated with a vulnerability. This motivates viewing JIT-VP as a broader problem encompassing multiple downstream tasks rather than solely the traditional binary classification formulation.

To address these gaps, we conduct a large-scale empirical study involving approximately 140,000 commits to investigate how different VFC and VIC labeling strategies affect the performance of JIT-VP techniques. Specifically, we consider four VFC labeling strategies: Manual labeling, precision-first labeling~\cite{hommersom2024automated}, quantity-first labeling~\cite{nguyen2023multi}, and an LLM-assisted approach~\cite{nguyen2025patchseeker}. We further consider four VIC labeling strategies: Manual labeling, Vulnerability-specific labeling~\cite{bao2022v}, Bug-generalization labeling~\cite{sliwerski2005changes}, and LLM-assisted labeling~\cite{tang2025llm4szz}. These strategies are among the most common methods to identify VFCs/VICs in JIT-VP (e.g.,~\cite{nguyen2025toward, nguyen2024code}) and related problems (e.g.,~\cite{cao2024snopy, wu2024vision}). We systematically evaluate the resulting combinations of VFC and VIC labels on datasets constructed from two large-scale open-source projects, OpenSSL and the Linux kernel. Each dataset configuration is evaluated using six representative JIT-VP models across three proposed downstream tasks: classification, prioritization, and localization. All experiments use the same curated benchmark set to ensure a consistent and fair comparison across labeling strategies. Specifically, for the localization task, we manually label the line-level ground truth and release the guidelines for future studies when labeling line-level vulnerability data.

Our study sheds light on how different VFC and VIC labeling strategies affect JIT-VP performance across downstream tasks. We find that classification and prioritization benefit most from labeling strategies that balance label quality and coverage, whereas highly accurate labels generated through human or LLM-assisted annotation show greater promise for line-level localization. We further identify the most effective dataset configuration for each task. In particular, manual VFC labeling combined with vulnerability-specific VIC labeling provides the most reliable overall performance for classification and prioritization, whereas precision-first VFC labeling combined with LLM-assisted VIC labeling performs best for localization. Notably, precision-first automated VFC labeling achieves performance comparable to manual labeling for both classification and prioritization, suggesting that high-quality automated VFC labeling can provide a scalable alternative to costly manual annotation. Traditional SZZ-based VIC labeling likewise remains competitive with manual labeling for prioritization, while LLM-assisted VIC labeling offers greater benefits for localization.

These findings suggest that there is no universally optimal labeling pipeline for JIT-VP. Instead, training data should be constructed according to the requirements of the intended downstream task, selecting labeling strategies that provide an appropriate balance among accuracy, coverage, scalability, and cost. More broadly, our results highlight task-oriented dataset construction as a practical direction for JIT-VP: scalable labeling can provide broad coverage for classification and prioritization, while more accurate human or LLM-assisted annotation can be selectively applied when fine-grained localization is required.

Our \textbf{key contributions} are as follows:
\begin{enumerate}
    \item We conduct a large-scale empirical study covering widely used data labeling approaches for JIT-VP in different tasks, providing insights into their impact on model performance and identifying promising directions for future research.

    \item We derive practical recommendations for selecting data labeling strategies to achieve competitive performance for different downstream JIT-VP tasks and objectives.

    \item We publicly release our datasets, guidelines, and replication package to promote open science and facilitate future research on JIT-VP. Our replication package is available at: \url{https://github.com/TheSyx04/Data-Labeling-for-JIT-VP}.
\end{enumerate}

The remainder of this paper is structured as follows. Section~\ref{sec:background} provides background on JIT-VP, downstream quality assurance tasks and labeling approaches. Section~\ref{sec:motivation} shows our motivation examples, while Section~\ref{sec:design} describes our study design, including research questions, data collection, studied models, experimental setups and implementation details. Section~\ref{sec:evaluation} presents our experimental results for each research question and Section~\ref{sec:qualitative} presents the qualitative analysis. Section~\ref{sec:discussion} discusses the implications of our findings and threats to validity, while Section~\ref{sec:related_work} shows related work. Finally, Section~\ref{sec:conclusion} concludes the paper.

\section{Background}
\label{sec:background}

\subsection{Just-In-Time Vulnerability Prediction: Is binary classification really enough?}

Software security defects are often introduced during routine code modifications and may remain undetected until they are discovered through testing, code review, or vulnerability reports. Detecting such defects after integration can be costly, particularly in projects with rapid development cycles and frequent code contributions. As a result, there is growing interest in techniques that can assess the security risk of code changes as early as possible, ideally when a change is submitted.

Just-In-Time Vulnerability Prediction (JIT-VP) has been proposed to address this challenge. Formally, JIT-VP is a commit-level problem that aims to provide developers with timely insights into the vulnerability risk of individual code changes using information available at commit time, such as source code modifications, commit messages, and development history instead of analyzing an entire codebase.

Typically, JIT-VP has been formulated as a binary classification problem. Given a commit, a JIT-VP model aims to determine whether the commit is likely to introduce a vulnerability into the software system and classify it as either vulnerability-introducing or not~\cite{perl2015vccfinder, nguyen2024code, nguyen2025toward}. However, as software systems continue to grow in scale and complexity while development cycles become increasingly rapid, the effort required to assure software security also increases. This growth highlights the need for approaches that account for the limited availability of human resources and help developers focus their attention on the most critical changes rather than investigate a large number of false alarms~\cite{yang2016effort}. Beyond identifying potentially vulnerable commits, developers can also benefit from knowing which changes are most likely to contain vulnerabilities and where the underlying vulnerability resides, together with explanations that facilitate investigation~\cite{fu2022linevul}. 
These practical requirements suggest that JIT-VP should not be viewed merely as a binary classification problem, but as a broader problem encompassing multiple tasks that support the software security assurance process.
% : is binary vulnerability prediction alone sufficient to effectively support JIT-VP in modern software development?

% Over the past decade, JIT-VP techniques have evolved from traditional machine learning models based on handcrafted process metrics, such as code churn, developer experience, and change history~\cite{kamei2012large,perl2015vccfinder}, to deep learning and ensemble-based approaches that exploit commit messages and code changes~\cite{hoang2019deepjit,zhou2022simple}. More recent studies further incorporate code-centric or graph-based representations to capture vulnerability-related patterns from source code changes~\cite{nguyen2024codejit}. 

% As modeling techniques become increasingly diverse, it is equally important to examine whether their reported improvements stem from advances in model design or from differences in the quality and construction of the training data. Despite the importance of data construction to model performance, systematic studies of how JIT-VP training data affect the downstream performances remain limited, leaving a gap in the literature.

\subsection{Formulating Quality Assurance Tasks for JIT-VP}

% JIT-VP aims to identify potentially vulnerable changes at the time they are introduced into a software system. 
As discussed in the previous subsection, JIT-VP should be viewed as a broader problem encompassing multiple tasks that support the software security assurance process. Yet, existing studies have primarily focused on the classification task and the predictive capability of JIT-VP models (e.g.,~\cite{perl2015vccfinder, nguyen2024code, nguyen2025toward}), while other aspects of practical vulnerability detection remain under-explored.
To provide a broader perspective, we consider three downstream tasks in our study: \textbf{classification}, \textbf{prioritization}, and \textbf{localization}. 

% Classification~\cite{perl2015vccfinder, nguyen2024code} follows the conventional JIT-VP formulation and evaluates whether a commit is likely to introduce a vulnerability. Prioritization~\cite{yang2016effort} evaluates whether JIT-VP models can effectively rank potentially vulnerable commits to help developers identify the most critical ones under limited inspection effort. Localization~\cite{fu2022linevul} further examines whether models can identify the specific code lines associated with a vulnerability, thereby providing more actionable information to support developers during investigation. For each task, we introduce its general objective and problem formulation.
% Depending on the intended use and evaluation setting, existing JIT-VP approaches can generally be formulated as classification~\cite{perl2015vccfinder}, prioritization~\cite{yang2016effort}, or localization~\cite{fu2022linevul} tasks. 
% Although classification has been extensively studied in the JIT-VP literature, prioritization and localization remain comparatively underexplored. 

% \paragraph{Classification.}
\textbf{Classification} is the most widely studied task in JIT-VP. Given an individual commit and its associated features, a JIT-VP model predicts whether the commit is likely to introduce a vulnerability. This task is typically formulated as a binary classification problem, where each commit is assigned a label indicating whether it is vulnerability-introducing or not. While this formulation provides a straightforward means of evaluating predictive performance, it implicitly treats all commits independently and does not explicitly account for the limited effort available to developers when inspecting model predictions.

% \paragraph{Prioritization.}
% In practice, developers often need to inspect a large number of commits while having limited time and resources. 
\textbf{Prioritization} aims to optimize developer effort by directing developers toward the most critical commits for inspection. This task can be formulated as a ranking problem under an effort constraint. Given a set of commits, a JIT-VP model assigns each commit a vulnerability score and ranks the commits according to their predicted likelihood of being vulnerability-introducing. The resulting ranking is then evaluated under predefined effort budgets to assess how effectively the model can identify vulnerable commits within a limited amount of developer effort. Compared with binary classification, this formulation better reflects the practical scenario in which developers can inspect only a fraction of the predicted commits and therefore need to focus their effort on the highest priority ones.
% The objective is to identify as many vulnerable commits as possible while minimizing the amount of code or changes that developers need to inspect. 
% , such as the Vulnerability Detection Ratio (VDR) under a given Marked Commit Ratio (MCR) or other inspection-budget settings.

% \paragraph{Localization.}
\textbf{Localization} aims to provide more actionable insights beyond predicting whether a commit is vulnerable. We formulate this task as a line-level localization problem. Given a model prediction for a potentially vulnerable commit, we derive an importance score for each input token and aggregate these scores to obtain the vulnerability importance of each changed line. The lines are then ranked according to their importance scores, with the goal of placing vulnerability-relevant lines near the top of the ranking. Such localization can help developers focus their inspection on the most relevant parts of a commit and reduce the effort required to identify potentially vulnerable changes. This task is particularly important for the practical usefulness of JIT-VP models, as developers ultimately need to inspect and validate the reported changes before taking corrective actions.

% Overall, these tasks represent complementary perspectives on JIT-VP. Classification focuses on predictive correctness, prioritization considers the practical constraints of developer effort, and localization aims to support the human interpretation and validation of model predictions. Despite that, current JIT-VP researches often focused soley on classification task (eg.,~\cite{perl2015vccfinder, nguyen2024codejit, nguyen2025toward}) while limited efforts were spent on these other two formulations.

\subsection{Data Labeling Approaches for JIT-VP}
While many JIT-VP studies focus on improving model architectures, the effectiveness of JIT-VP also depends heavily on the quality and construction of the underlying datasets. In vulnerability studies, data labeling is particularly challenging because vulnerable commits are not directly observable. A common practice is to first identify vulnerability-fixing commits (VFCs), which are commits that patch known vulnerabilities, and then trace back from these fixes to locate vulnerability-introducing commits (VICs), the actual causes of the problem. The resulting VICs are treated as vulnerable instances, while other commits, including VFCs and vulnerability-neutral commits (VNCs), which are the commits unrelated to any existing vulnerabilities, are treated as safe instances during model training and evaluation. 

However, constructing reliable VFC and VIC labels remains difficult because each labeling strategy captures different aspects of vulnerability-related commits. For VFCs, manually curated vulnerability reports~\cite{OpenSSLPatches, LinuxCVE} and official security advisories~\cite{NVD} generally provide high-confidence labels, but they require substantial human effort and may cover only publicly documented vulnerabilities. Automated VFC labeling techniques~\cite{hommersom2024automated, nguyen2023multi} with fixed or flexible criteria and rules can scale to larger repositories and uncover more candidate fixes, but they often depend on predefined patterns in commit messages or vulnerability records, making them sensitive to missing, ambiguous, or inconsistent textual evidence. More recent learning-based and LLM-assisted approaches~\cite{nguyen2025patchseeker} can exploit richer semantic information from vulnerability descriptions and code changes, but they may introduce additional computational cost and uncertainty in their predictions. Similarly, VICs are commonly identified using different variants of the SZZ algorithm~\cite{bao2022v, sliwerski2005changes}, developer-provided metadata~\cite{LinuxKernelPatches, rosa2021evaluating}, or LLM-assisted~\cite{tang2025llm4szz} approaches. Typically, SZZ-based methods~\cite{bao2022v, sliwerski2005changes} are scalable and widely adopted, but they may misattribute the root cause of a vulnerability due to noisy blame information or complex code histories. Developer-oracle data~\cite{LinuxKernelPatches} can be more precise when available, but it is project-specific and often incomplete. LLM-assisted methods may improve contextual reasoning over code changes, but their reliability and cost can vary across projects. These differences suggest that labeling strategies are not interchangeable; they may produce datasets with different levels of coverage, noise, and bias. Therefore, it is necessary to systematically compare their downstream impact on JIT-VP performance. However, this domain remains under-explored in the current literature.

\begin{figure}
    \centering
    \includegraphics[width=0.5\linewidth]{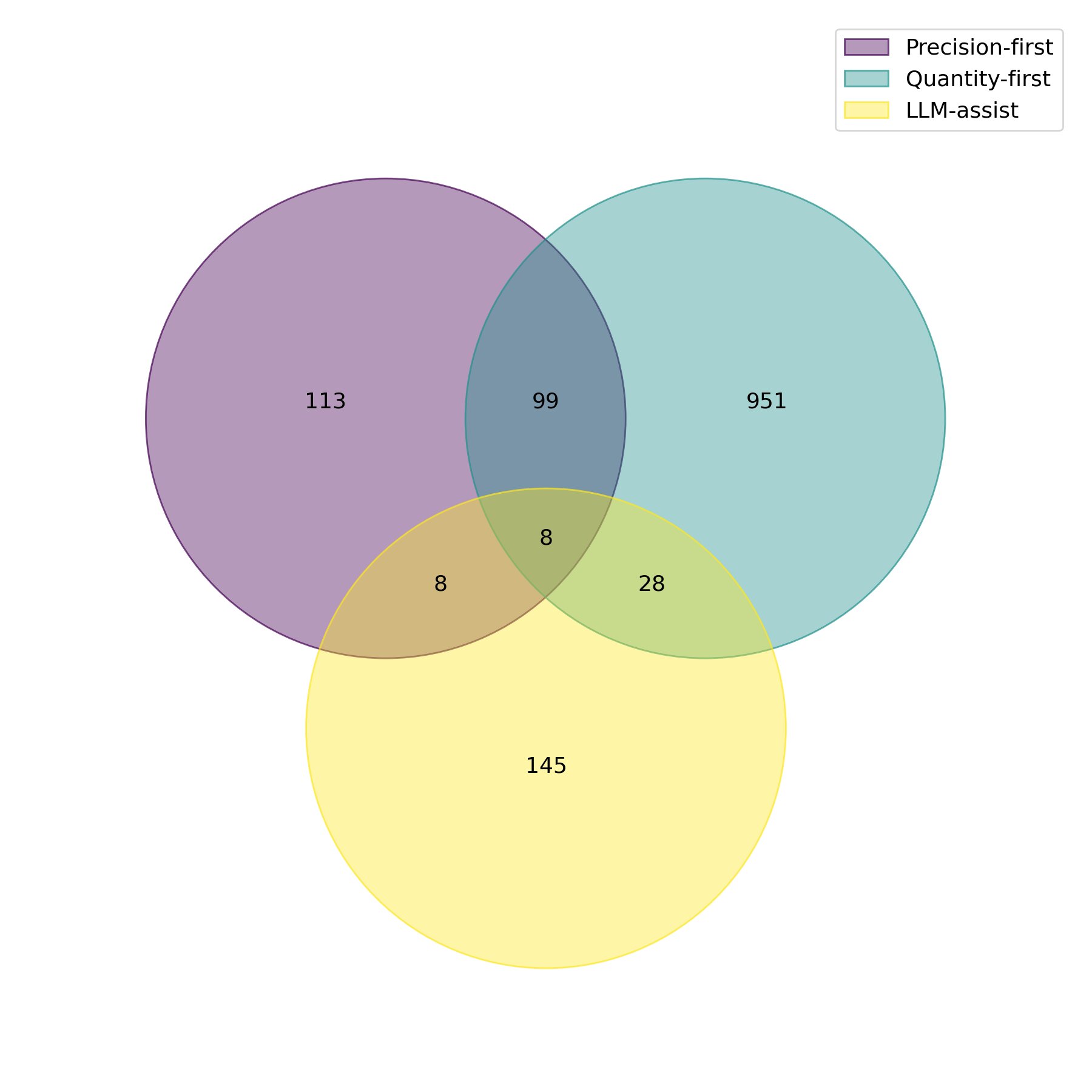}
    \caption{VFCs identified in the OpenSSL project by three automated methods.}
    \label{fig:motivation_1}
\end{figure}

\section{Motivation}
\label{sec:motivation}

Typically, constructing a JIT-VP dataset involves two stages: first identifying vulnerability-fixing commits \textbf{(VFCs)}, and then tracing them back to the vulnerability-introducing commits \textbf{(VICs)}. As discussed in Section~\ref{sec:background}, various strategies have been proposed for both labeling tasks. These strategies, however, involve different trade-offs among accuracy, scalability, and cost. This issue is particularly pronounced for automated labeling methods, where differences in their underlying assumptions can lead to substantially different labels.

To illustrate this phenomenon, we conducted a preliminary study comparing three automated approaches for identifying VFCs: Precision-first labeling~\cite{sap_project_kb}, Quantity-first labeling~\cite{nguyen2023multi}, and LLM-assist labeling~\cite{nguyen2025patchseeker}. Figure~\ref{fig:motivation_1} presents the resulting VFC labels. Although Precision-first labeling and LLM-assisted labeling operate on the same input information, they identify substantially different sets of VFCs, sharing only 16 commits, while their respective unique sets contain 113 and 145 commits. In contrast, the Quantity-first labeling, which identifies candidate VFCs by matching vulnerability-related regular expressions against commit messages, identifies the largest number of VFCs, with 1,086 commits, of which 951 are unique to this method. These results suggest that seemingly reasonable labeling strategies can produce substantially different VFC datasets.

\begin{figure}
    \centering
    \includegraphics[width=0.7\linewidth]{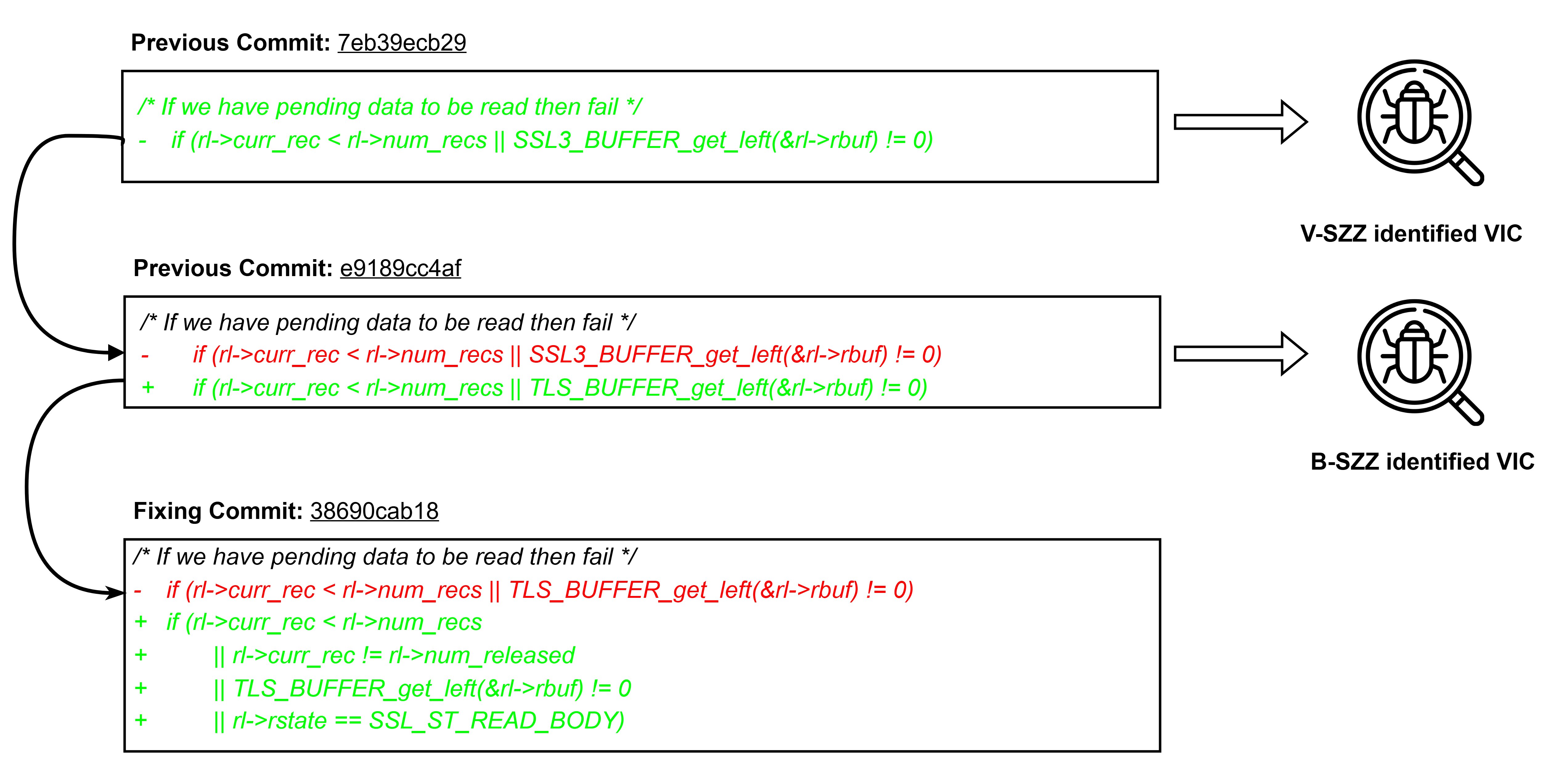}
    \caption{VICs inferred from VFC 38690cab18 in the OpenSSL project labeled by two strategies: V-SZZ and B-SZZ}
    \label{fig:motivation_2}
\end{figure}

The choice of VIC identification strategy introduces further variation. Figure~\ref{fig:motivation_2} illustrates the VIC labels obtained for the VFC \texttt{38690ca}\footnote{\url{https://github.com/openssl/openssl/commit/38690ca}} associated with CVE-2024-4741~\cite{cve-2024-4741} in OpenSSL~\cite{openssl}, using two SZZ-based strategies: V-SZZ~\cite{bao2022v}, which is specifically designed for vulnerability analysis, and B-SZZ~\cite{sliwerski2005changes}, which targets general bug-introducing changes. Although both belong to the SZZ family, they differ in how they trace a fixing change to its historical origin. V-SZZ recursively traces the affected code to identify vulnerability-introducing changes and identifies \texttt{7eb39ec}\footnote{\url{https://github.com/openssl/openssl/commit/7eb39ec}} as the VIC, whereas B-SZZ traces the affected lines to their preceding revisions and identifies \texttt{e9189cc}\footnote{\url{https://github.com/openssl/openssl/commit/e9189cc}} instead. The two strategies also differ in computational cost: V-SZZ performs deeper historical tracing and can therefore require substantially more computation and memory, whereas B-SZZ performs shallower tracing and is generally more efficient.

% Neither strategy is universally preferable, rather each embodies 
These variations arise from different assumptions about how vulnerability-related patches and their introducing changes should be identified. Although different approaches may report state-of-the-art predictive performance on particular benchmarks (e.g.,~\cite{nguyen2025mapping, tang2025llm4szz, bao2022vszz}), they inherently entail different trade-offs among labeling specificity, computational cost, and coverage. For example, a manually curated and highly selective dataset may provide accurate labels but fail to fully capture the characteristics of diverse software projects, whereas an LLM-based approach may incur substantially higher computational costs than simple heuristic methods. Such differences in the underlying labeling process can ultimately affect the resulting training data and, consequently, the conclusions drawn about model performance. A labeling strategy that prioritizes coverage may introduce noisy labels that degrade model generalization, whereas a highly selective strategy may provide cleaner labels at the cost of insufficient coverage of the underlying data distribution~\cite{song2022learning, hendrycks2018using}. As a result, neither labeling accuracy nor quantity alone is sufficient to characterize the suitability of a labeling strategy. Despite its importance, however, relatively little work has systematically investigated how alternative labeling strategies influence downstream JIT-VP performance.

Meanwhile, existing JIT-VP studies also employ different labeling pipelines, making their results difficult to interpret independently of the underlying label construction process. For example, Perl et al.~\cite{perl2015vccfinder} identified VFCs using a combination of commits explicitly referenced by CVE records and commits containing CVE references in their messages, and identified VICs using a customized SZZ algorithm that traced deleted lines as well as their neighboring lines. Nguyen et al.~\cite{nguyen2024code} reused VFCs from prior work and identified VICs by tracing semantically related lines through control dependencies. Nguyen et al.~\cite{nguyen2025toward} combined CVE-referenced and researcher-labeled VFCs and identified VICs using developer-informed references and V-SZZ. These differences illustrate that JIT-VP datasets are constructed using substantially different assumptions and labeling procedures, yet the extent to which these choices influence downstream model performance remains unclear.

Motivated by these observations, this study adopts a data-centric perspective on JIT-VP to systematically and uniformly investigate how different VFC and VIC labeling strategies, reflecting commonly used labeling practices, affect downstream JIT-VP performance. By comparing manual labels, automated labels, and their combinations across formulations of the JIT-VP problem, we aim to understand which labeling configurations provide effective training data for realistic and generalizable JIT-VP systems for practical use in the software quality assurance process.

% Another observation from our study is that different JIT-VP tasks require different levels of label granularity. Classification and vulnerability prioritization can be formulated using binary commit-level labels, whereas line-level localization requires finer-grained annotations that identify the vulnerable lines within a commit. However, as discussed in Section~\ref{sec:background}, relatively few studies have investigated vulnerability prioritization and localization in the context of JIT-VP. Consequently, existing JIT-VP datasets have largely been constructed with commit-level classification as the primary objective, while datasets supporting other downstream tasks remain limited despite their practical importance and challenges.

% Consider CVE-2021-28039~\cite{CVE-2021-28039} in the Linux kernel as an example. Its VFC (commit 8822139\footnote{\url{https://github.com/torvalds/linux/commit/8822139}}) contains 21 added and 26 deleted lines, while its VIC (commit 9e2369c\footnote{\url{https://github.com/torvalds/linux/commit/9e2369c}}) contains 219 added and 15 deleted lines. When combined with the complex dependencies and interactions involved in vulnerability-inducing changes, such large commits illustrate the potential limitations of representing vulnerability information solely as a binary label. A positive label indicates that a commit is associated with a vulnerability, but does not reveal which parts of the change are relevant to the vulnerability or where developers should focus their attention.

\begin{figure}
    \centering
    \includegraphics[width=\linewidth]{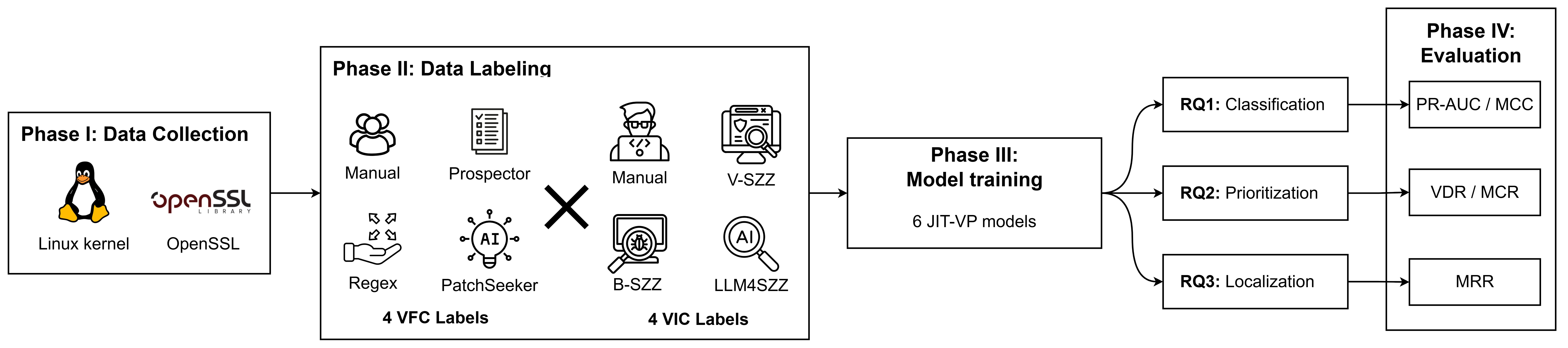}
    \caption{Overview of the study workflow, including four main phases: Data Collection, Data Labeling, Model Training, and Evaluation.}
    \label{fig:flow}
\end{figure}

\section{Study Design}
\label{sec:design}

In this section, we introduce our research questions and study scope, including the labeling methods and models investigated, problem formulations, experimental setup, and implementation details. Figure \ref{fig:flow} provides an overview of the workflow adopted in this study, which consists of four main stages: data collection, data labeling, model training, and evaluation.

\subsection{Research Questions}
This study investigates how training data constructed using different labeling strategies affects the performance of JIT-VP models. Specifically, we examine this impact across three scenarios: classification, prioritization, and localization. Answering these questions enables us to understand whether the choice of labeling strategy has task-dependent effects and to identify the
strategies that are most suitable for different practical JIT-VP scenarios. Accordingly, we formulate the following research questions:

\textbf{RQ1}. \textit{How do different labeling strategies affect the performance of JIT-VP models in the classification task?}

\textbf{RQ2}. \textit{How do different labeling strategies affect the performance of JIT-VP models in the prioritization task?}

\textbf{RQ3}. \textit{How do different labeling strategies affect the performance of JIT-VP models in the localization task?} 

\subsection{Data Collection}
\label{sec:design-data}
In this subsection, we describe our data collection methodology.

\subsubsection{Commit Curation}
\begin{table}[t]
\centering
\caption{Distribution of Vulnerability Fix Durations in the Linux kernel.}
\label{tab:linux_vul}
\begin{tabular}{lr}
\hline
\textbf{Fix Duration} & \textbf{Percentage (\%)} \\
\hline
$< 1$ week           & 0.25  \\
1 week -- 1 month    & 2.51  \\
1 month -- 1 year    & 29.20 \\
1 year -- 5 years    & 40.27 \\
$> 5$ years          & 27.77 \\
\hline
\textbf{Total}       & \textbf{100.00} \\
\hline
\end{tabular}
\end{table}

To ensure that our evaluation reflects real-world development practices, we collected commit data from two widely used open-source projects: OpenSSL~\cite{openssl} and the Linux kernel~\cite{linux}. These projects were selected for two reasons. First, both are large-scale, influential systems that are representative of modern software development environments. Second, they are actively maintained by experienced security practitioners and provide high-quality vulnerability information, which is essential for constructing reliable datasets.

However, the Linux kernel contains over one million commits, reflecting the cumulative contributions of open-source developers worldwide. As our study involved multiple data labeling techniques and extensive model training, directly processing the entire commit history would impose substantial computational and resource demands. We therefore aimed to reduce the size of the Linux dataset to enable efficient experimentation within our available computational resources while retaining a sufficiently large and representative sample to provide reasonable confidence in the resulting vulnerability labels.

To determine an appropriate temporal window, we first analyzed the time required to identify and fix vulnerabilities in the Linux kernel. Using the root-cause commits referenced by Linux kernel vulnerability-fixing patches reported in the NVD~\cite{LinuxKernelPatches, lyu2024evaluating}, we measured the time between vulnerability introduction and its corresponding fix. As shown in Table~\ref{tab:linux_vul}, approximately 72\% of the vulnerabilities in our analysis were fixed within five years of their introduction. Based on this observation, we restricted the Linux dataset to commits made between 2020 and 2021, thereby allowing vulnerabilities introduced during this period at least five years to be identified and fixed before our data collection cut-off date.

This temporal restriction reduced the Linux dataset from more than one million commits to approximately 120,000 commits, substantially reducing the computational cost of our large-scale experiments while retaining a sufficiently large dataset for model training. However, the restriction may exclude vulnerabilities whose fixes were identified more than five years after their introduction. We therefore interpret the five-year window as an empirical coverage criterion rather than a guarantee of complete vulnerability coverage. Under the assumption that vulnerabilities are typically identified and fixed within five years, the resulting dataset is expected to capture a substantial proportion of vulnerabilities introduced during the study period. This choice is further supported by prior empirical studies of vulnerability lifetime~\cite{277196, iannone2022secret}. For example, Alexopoulos et al.~\cite{277196} investigated the time vulnerabilities remain in code before being fixed and reported an average lifetime of approximately four years. For the Linux kernel specifically, they reported mean and median lifetimes of 1,732.97 and 1,363.5 days, respectively. Overall, the selected temporal window provides a practical balance between computational feasibility and vulnerability coverage.

We collected all commits made to OpenSSL prior to 11 October 2025 and all commits to the Linux kernel between 1 January 2020 and 31 December 2021. We then filtered the collected commits to exclude changes unrelated to the projects' primary development activities, such as merge commits and blank commits, thereby improving the relevance of the resulting datasets for vulnerability prediction. The remaining commits were ordered chronologically to preserve their temporal structure and prevent data leakage, since JIT-VP uses historical commits to predict vulnerabilities in future commits. This process resulted in 24,787 commits for OpenSSL and 115,471 commits for Linux, which were subsequently used to construct the chronological training, validation, and test sets.

\subsubsection{Data Configuration Constructions}

\begin{table*}[t]
  \centering
  \caption{Dataset statistics for OpenSSL and the Linux kernel under different configurations.
           VIC = vulnerability-introducing commits,
           VFC = vulnerability-fixing commits,
           VNC = vulnerability-neutral commits.}
  \label{tab:design_data_stat}
  \setlength{\tabcolsep}{5pt}

  \begin{minipage}[t]{0.44\linewidth}
    \centering
    \small
    \textbf{OpenSSL}\\[4pt]
    \begin{tabular}{l r r r}
      \toprule
      \textbf{Data} & \textbf{VIC} & \textbf{VFC} & \textbf{VNC} \\
      \midrule
      0-1 &  144 &  114 & 17{,}303 \\
      0-2 &  105 &  114 & 17{,}317 \\
      0-3 &   30 &  114 & 17{,}348 \\
      \midrule
      1-1 &  387 &  274 & 15{,}592 \\
      1-2 &  521 &  274 & 14{,}971 \\
      1-3 &   60 &  274 & 17{,}140 \\
      \midrule
      2-1 & 1{,}348 & 1{,}086 & 15{,}073 \\
      2-2 & 1{,}893 & 1{,}086 & 14{,}737 \\
      2-3 &     287 & 1{,}086 & 17{,}161 \\
      \midrule
      3-1 &  378 &  189 & 15{,}664 \\
      3-2 &  409 &  189 & 15{,}463 \\
      3-3 &   70 &  189 & 16{,}457 \\
      \bottomrule
    \end{tabular}
  \end{minipage}
  \hfill
  \begin{minipage}[t]{0.54\linewidth}
    \centering
    \small
    \textbf{Linux kernel}\\[4pt]
    \begin{tabular}{l r r r}
      \toprule
      \textbf{Data} & \textbf{VIC} & \textbf{VFC} & \textbf{VNC} \\
      \midrule
      0-0        &    156 &    648 & 76{,}193 \\
      0-1        &    245 &    648 & 76{,}104 \\
      0-2        &    316 &    648 & 76{,}033 \\
      0-3        &    174 &    648 & 76{,}175 \\
      \midrule
      1-0        &    158 &    689 & 76{,}191 \\
      1-1        &    246 &    689 & 76{,}103 \\
      1-2        &    317 &    689 & 76{,}032 \\
      1-3        &    177 &    689 & 76{,}172 \\
      \midrule
      2-0        &      602 &  8{,}749 & 75{,}747 \\
      2-1        &  2{,}234 &  8{,}749 & 74{,}115 \\
      2-2        &  3{,}694 &  8{,}749 & 72{,}655 \\
      2-3        &      895 &  8{,}749 & 75{,}454 \\
      \midrule
      3-0        &    196 &    862 & 76{,}153 \\
      3-1        &    358 &    862 & 75{,}991 \\
      3-2        &    475 &    862 & 75{,}874 \\
      3-3        &    230 &    862 & 76{,}119 \\
      \bottomrule
    \end{tabular}
  \end{minipage}
\end{table*}

As discussed in Section~\ref{sec:background}, constructing training data for JIT-VP is inherently challenging because vulnerability-introducing commits (VICs) are elusive and are typically identified by tracing them back from vulnerability-fixing commits (VFCs). Consequently, each JIT-VP dataset configuration is defined by a pair of VFC and VIC labeling strategies, with different combinations producing different training datasets.

% Pre study: number of percentage of labels 1/0 in diff sets
In this study, we employed multiple VFC and VIC labeling strategies to systematically investigate the impact of labeling techniques on the performance of JIT-VP models. Each data configuration is denoted as dataset$_{x-y}$, involving the use of method VFC$_x$ and VIC$_y$, where $x,y \in \{0,1,2,3\}$.

The VFC and VIC labeling strategies investigated in this study are as follows:
% Each labeling approach involves different trade-offs between accuracy, scalability, and labeling cost.

\textbf{Vulnerability-Fixing Commits (VFCs).}
\begin{itemize}[leftmargin=*, label={}]
    \item \textbf{VFC$_{0}$ - \textit{Manual Label}:} Manual labels are derived through human effort, typically from researchers' manual analysis or developers' security reports. This approach generally provides high labeling accuracy but requires substantial manual effort and is difficult to scale. In this study, we obtained manual VFC$_0$ labels from official  security patches and vulnerability reports recorded in the National Vulnerability Database (NVD)~\cite{OpenSSLPatches, LinuxCVE}.

    \item \textbf{VFC$_{1}$ - \textit{Precision-first Label}:} Precision-first labeling aims to expand manually curated data while maintaining high label precision. These approaches typically employ rule-based techniques with strict criteria, trading scalability for precision. One example is Prospector~\cite{hommersom2024automated}, a rule-based approach for linking NVD records to their corresponding VFCs. Given an NVD entry, Prospector scans the commit history to identify candidate commits and ranks their relevance based on heuristic rules, including checking whether a commit is explicitly referenced in the NVD record and matching security-relevant keywords in commit messages and code diffs. By prioritizing strict matching criteria over scalability, Prospector represents a precision-first strategy in this study. To reproduce this labeling strategy, we used the authors' official package~\cite{sap_project_kb} to generate the VFC$_1$ labels.
    % Although this approach generally produces high-quality labels, it tends to identify a relatively small number of VFCs.

    \item \textbf{VFC$_{2}$ - \textit{Quantity-first Label}:} Quantity-first approaches aim to broaden the coverage of VFC labels, potentially at the expense of labeling accuracy. They prioritize identifying a large and diverse set of candidate VFCs, while relying on noise-tolerant models or learning techniques to mitigate the resulting label noise. These approaches provide a fast, inexpensive, and highly scalable means of generating large numbers of VFC labels. In this study, the representative quantity-first approach employs regular expressions constructed from vulnerability-related keywords, as proposed by Nguyen et al.~\cite{nguyen2023multi}, to filter commit messages and identify candidate VFCs. We selected this approach to represent the quantity-first strategy because of its high scalability, low computational cost, and ease of reproduction.
    % However, this increased coverage typically comes at the cost of low precision.

    \item \textbf{VFC$_{3}$ - \textit{LLM-assisted Label}:} LLM-assisted approaches aim to balance labeling accuracy and scalability while improving label explainability through the semantic understanding capabilities of large language models (LLMs). Rather than relying solely on predefined rules or keyword matching, these approaches can interpret the intent and context of vulnerability reports and security patches to establish links between them and their corresponding VFCs. However, leveraging LLMs typically incurs substantially higher computational and monetary costs than rule-based alternatives. We selected PatchSeeker as a representative approach~\cite{nguyen2025patchseeker} for this strategy. PatchSeeker leverages LLMs to rank candidate VFCs based on information extracted from NVD records. In this study, we followed the experimental settings and configuration of the original PatchSeeker study~\cite{nguyen2025patchseeker_replication} to generate the VFC$_3$ labels.
\end{itemize}

\textbf{Vulnerability-Introducing Commits (VICs).}
\begin{itemize}[leftmargin=*, label={}]
    \item \textbf{VIC$_{0}$ - \textit{Manual Label}:} Manual labeling aims to achieve the highest possible labeling accuracy at the cost of substantial human effort. Consequently, manually curated datasets are typically highly accurate but limited in size. Moreover, manual labeling practices can be project-specific, as different organizations may follow different policies for documenting and managing vulnerabilities. For example, the Linux kernel recommends that a vulnerability-fixing patch explicitly reference the commit being fixed in its commit message~\cite{LinuxKernelPatches, lyu2024evaluating}, whereas no equivalent practice is adopted by OpenSSL. In this study, we therefore obtained manual VIC labels only for the Linux kernel by extracting these explicit references from VFC commit messages. This practice allows us to identify VICs with high confidence while avoiding additional manual effort.

    \item \textbf{VIC$_{1}$ - \textit{Vulnerability-specific Label}:} Manually identifying vulnerability-introducing commits requires substantial human effort, thus motivating automated approaches to act as a more scalable alternative. These approaches exploit vulnerability-specific information to trace VFCs back to their corresponding VICs. One widely used approach is V-SZZ~\cite{bao2022v}, a variant of the SZZ algorithm specifically designed to identify VICs from known VFCs. V-SZZ is based on the observation that the commit introducing a vulnerable change can often be identified by tracing the vulnerable lines modified by a fixing patch back through the repository history. Given a VFC, V-SZZ identifies the lines changed by the fixing patch and uses repository history to trace these lines to the commits that originally introduced them, thereby identifying candidate VICs. In this study, we obtained the VIC$_{1}$ labels by utilizing the original replication package~\cite{bao2022vszz}.
    
    \item \textbf{VIC${2}$ - \textit{Bug-generalization Label}:} Vulnerabilities can be viewed as a specific type of software defect, reflecting the inherent relationship between vulnerabilities and bugs. Consequently, defect identification techniques can provide a scalable means of identifying candidate VICs without over-relying on vulnerability-specific rules, making them more generalizable across projects while potentially introducing additional label noise. B-SZZ~\cite{sliwerski2005changes} is the original SZZ algorithm for identifying bug-introducing commits. We selected B-SZZ as the representative of this strategy because it remains one of the most widely used and influential SZZ variants despite the subsequent development of numerous extensions. Given a vulnerability-fixing patch, B-SZZ applies line-level blame to trace deleted or modified lines in the patch back to their previous revisions and identifies the corresponding commits as bug-introducing commits. In our study, we obtained the VIC${_2}$ labels using the replication package from the empirical evaluation of different SZZ variants by Rosa et al.~\cite{rosa2021szzreplication}.
        
    \item \textbf{VIC${3}$ - \textit{LLM-assisted Label}:} LLM-assisted approaches aim to improve the scalability and accuracy of VIC labeling by leveraging the semantic understanding of large language models (LLMs) to incorporate commit intent and contextual information. Compared with traditional SZZ-based techniques, these approaches can exploit richer semantic information when identifying vulnerability-introducing commits. However, they typically incur substantially higher computational and monetary costs due to the complexity of LLM-based processing. We selected LLM4SZZ~\cite{tang2025llm4szz} as a representative of this strategy because it is a promising approach that leverages LLMs to improve VIC identification beyond traditional SZZ-based techniques. We obtained the VIC${_3}$ labels using the original replication package~\cite{Tang2025}.
    \end{itemize}

To construct the training datasets, we used the manually curated labels (VFC$_0$ and VIC$_0$) as the base configuration. This means that the VFCs and VICs identified through manual labeling were included in every dataset. Starting from this common base, we constructed different dataset configurations by incorporating VFCs and VICs identified using alternative labeling strategies. This design allowed us to compare different combinations of VFC and VIC labeling strategies as complementary sources of labels that can extend the coverage of manually curated data while retaining its high accuracy and reducing the substantial human effort required for manual labeling.

For each dataset$_{x-y}$, we first identified VFCs using the corresponding VFC labeling strategy VFC$_x$ and combined them with the manually curated VFCs (VFC$_0$). We then traced these VFCs to their corresponding VICs using the VIC labeling strategy VIC$_y$. The newly identified VICs were combined with the manually curated VICs (VIC$_0$), and all resulting VICs were labeled as positive instances, while all remaining commits were treated as negative instances. 

For each configuration, we chronologically partitioned the commits into training, validation, and test sets using a 75\%/5\%/20\% split, respectively. This chronological split prevents information from future commits from leaking into model training and reflects the temporal nature of JIT-VP, in which models are expected to predict the vulnerability risk of newly arriving commits. To further prevent data leakage, we removed from each training set all commits that appeared in the common test set, which are discussed in Subsection~\ref{sec:benchmark}. Finally, Table~\ref{tab:design_data_stat} summarizes the class distributions of the resulting training dataset configurations.

\subsubsection{Data Benchmarks}
\label{sec:benchmark}
To ensure a fair and consistent comparison across experimental settings, we used a common benchmark test set for all experiments within each project. 

For RQ1 and RQ2, we used the test set of data$_{0-0}$ for Linux and the test set of data$_{0-1}$ for OpenSSL as the common evaluation benchmarks. These configurations were selected because they provided the most reliable available VFC and VIC labels for their respective projects. In particular, the Linux kernel has both manually labeled VFC$_0$ and VIC$_0$, while OpenSSL does not have VIC$_0$ because its development practice does not provide developer-informed references to the corresponding vulnerable commits. Hence, V-SZZ was selected as the alternative option since V-SZZ is a vulnerability-specific SZZ variant that has been widely used for vulnerability-introducing commit identification~\cite{bao2022v} and has been used in the literature as the alternative source of ground truth when a human-validated source is not available~\cite{nguyen2025toward}. Using these configurations as common benchmarks therefore provided a high-confidence basis for evaluating the effects of alternative labeling strategies. To prevent data leakage, we removed from the training and validation sets of every configuration any commit that appears in the corresponding common test set. 

For RQ3, we required line-level ground truth in addition to commit-level vulnerability labels. Therefore, we manually annotated the vulnerable lines associated with the benchmark VICs to establish reliable ground truth for evaluating vulnerability localization. We first discussed the annotation criteria and developed a common annotation guideline. The annotation was then conducted in two sequential stages. In the first stage, the first annotator identified the vulnerable lines for each benchmark commit according to the annotation guideline. In the second stage, the second annotator reviewed the annotations and assessed the vulnerable lines for the same commits. The annotation involved two security researchers, each with at least two years of experience working with vulnerabilities in open-source projects. After completing the two stages, the annotators compared their results and discussed any disagreements to reach a consensus. We then measured the agreement between the annotations before consensus resolution using Cohen's kappa~\cite{cohen1960coefficient} to assess the consistency of the annotation process.

For the line-level vulnerability data annotation guideline, we first examined all VFCs that could be linked to an NVD/CVE record to ensure that the resulting line-level annotations were associated with known vulnerabilities. We then identified the corresponding VICs using the most reliable available source for each project. For Linux, we leveraged developer-provided references in commit messages~\cite{LinuxKernelPatches}, as Linux vulnerability fixes are often accompanied by explicit references to the commits that introduced the vulnerability. For OpenSSL, where such references are generally unavailable, we used V-SZZ to trace vulnerability-fixing changes back to candidate VICs~\cite{bao2022v} for the same reason that we chose this strategy to construct the common test set in RQ1 and RQ2. These project-specific strategies were selected to maximize the reliability of the VIC identification process.

For each VIC and VFC pair, we then manually examined the changes between the vulnerable and patch commits and identified the common code changes associated with the vulnerability. We excluded comments and changes to non-code files, such as text and Markdown files. We considered a line of code to be vulnerable if the same line was subsequently deleted or modified by the corresponding fixing commit. This criterion follows the underlying intuition of SZZ-based approaches, which trace fixing changes back to the code that introduced them~\cite{sliwerski2005changes}. While this criterion may not capture vulnerabilities involving inter-procedural dependencies~\cite{li2024effectiveness} or fixes that only add new code~\cite{rezk2021ghost}, we adopted this conservative criterion to prioritize annotation accuracy and consistency. We retained only commits for which the vulnerable lines could be identified with sufficient confidence and agreement between the annotators. Commits for which the vulnerable lines could not be reliably determined using these criteria were excluded from the evaluation. Finally, we removed any benchmark commits that appeared in the training data to prevent data leakage. 

Overall, this process resulted in \textbf{28} Linux commits and \textbf{12} OpenSSL commits with a total of \textbf{161} manually validated vulnerable lines out of \textbf{33,480} lines of code. The annotation process cost approximately \textbf{80} man-hours and achieved a Cohen's Kappa of \textbf{0.80}, indicating substantial agreement between the two annotation stages. These commits and their corresponding vulnerable lines constituted the ground truth used to evaluate the explanation capability in RQ3.

\subsection{Studied Models}
\label{sec:models}

In this subsection, we describe the JIT-VP models included in our study. Our selection covered diverse learning paradigms commonly adopted in commit-level prediction, including traditional machine learning (ML), ensemble learning, deep learning (DL), and pre-trained language models. This diversity enabled us to examine whether the impact of labeling strategies is consistent across different model architectures and feature representations. We selected six representative models: three ML models, namely, Logistic Regression (LR), LAPredict, and TLEL; two DL models, namely, DeepJIT and SimCom, and a pre-trained language model JITFine.

\textbf{LR}~\cite{kamei2012large} represents a traditional feature-based approach to commit-level prediction. It uses manually engineered features proposed by Kamei et al.~\cite{kamei2012large}, which characterize multiple aspects of a commit, including \textit{diffusion}, \textit{size}, \textit{purpose}, \textit{history}, and \textit{experience}. The work of Kamei et al. established these expert features as a widely used baseline for both JIT-DP and subsequent JIT-VP research. We included LR as a simple and interpretable representative of conventional feature-based ML approaches.

\textbf{LAPredict}~\cite{zeng2021deep} is an intentionally simple prediction model based solely on Logistic Regression and a single feature: the number of lines added to a commit. The model is motivated by observations that lines added is among the most influential predictors within the set of expert features proposed by Kamei et al.~\cite{kamei2012large}. Despite its simplicity, LAPredict provides a competitive baseline for examining whether complex models consistently outperform lightweight approaches. 
We included LAPredict to assess how labeling strategies affect a model with minimal feature-engineering and modeling complexity.

\textbf{TLEL}~\cite{yang2017tlel} represents the traditional ML paradigm in our study. It is an ensemble-based approach built upon a two-layer learning architecture, where the first layer consists of multiple Random Forest classifiers and the second layer combines their predictions using bagged Decision Trees. Originally proposed for Just-In-Time Defect Prediction (JIT-DP), TLEL achieved state-of-the-art performance among ML-based approaches and demonstrated the effectiveness of ensemble learning for commit-level risk prediction. It was reported as the best-performing JIT-VP model by Lomio et al.~\cite{lomio2022just}. We included TLEL as a representative and competitive baseline for evaluating the capabilities of conventional ML techniques in the JIT-VP domain.

\textbf{DeepJIT}~\cite{hoang2019deepjit} is a pioneering deep learning model for commit-level prediction tasks. It employs two Convolutional Neural Networks (CNNs) to automatically learn representations from commit messages and code changes. The extracted features are concatenated and passed through fully connected layers to generate the final prediction. By eliminating the need for manual feature engineering, DeepJIT demonstrated the effectiveness of deep representation learning for software quality prediction and subsequently inspired many JIT vulnerability prediction approaches~\cite{le2020deep}. We included DeepJIT to represent the family of deep learning-based methods in JIT-VP.

\textbf{SimCom}~\cite{zhou2022simple} is included to represent hybrid ensemble approaches for commit-level prediction. The model combines two complementary components: \textit{Sim}, a lightweight decision tree trained on traditional commit change metrics~\cite{kamei2012large}, and \textit{Com}, a CNN-based model that learns semantic representations from commit data in a manner similar to DeepJIT~\cite{hoang2019deepjit}. Each component independently estimates the risk associated with a commit, and the final prediction is obtained by averaging their outputs. By integrating both handcrafted and learned features, SimCom demonstrates the potential of combining different models to improve prediction performance, making it a representative ensemble-learning approach for JIT-VP tasks.

\textbf{JITFine}~\cite{ni2022best} is a pre-trained language model-based approach for commit-level prediction. It combines traditional expert features with deep representations derived from CodeBERT~\cite{feng2020codebert}. Specifically, JITFine uses CodeBERT to encode commit messages and code changes and then integrates the resulting semantic representations with manually engineered commit features through an integrated feature learning network. The learned commit representation is subsequently used to predict whether the commit is defective. We included JITFine as a representative state-of-the-art approach based on pre-trained code language models, enabling us to investigate whether more expressive representations are more or less sensitive to the choice of training-data labeling strategy.

\subsection{Experimental Setups}
\label{sec:exp_design}
\label{sec:metrics}

In this section, we describe the experimental formulations, setups, and evaluation metrics of each downstream task, namely: classification, prioritization, and localization used to address RQ1, RQ2, and RQ3, respectively.

\subsubsection{\textbf{RQ1}} \textbf{Classification}

\textbf{Classification}~\cite{perl2015vccfinder, nguyen2025toward} aims to filter out the majority of non-vulnerable commits while retaining vulnerable ones for further analysis. To simulate this process, we formulated classification as a \emph{binary prediction task}, where commits with a predicted vulnerability probability of at least 0.5 are classified as vulnerable.
We evaluated classification performance using two metrics: Matthews Correlation Coefficient (\textbf{MCC}) and Area Under the Precision--Recall Curve (\textbf{PR-AUC}), with PR-AUC as the primary metric used for checkpoint selection. 

\textbf{MCC} is a special case of the $\phi$ coefficient~\cite{matthews1975comparison} and evaluates binary classification based on the number of True Positives (TP), True Negatives (TN), False Positives (FP), and False Negatives (FN). Unlike accuracy, MCC accounts for all four outcomes and has been shown to provide a reliable measure of classification performance under severe class imbalance~\cite{chicco2020advantages, zhu2020performance, le2022use}. MCC is formally defined as:
\begin{equation*}
    MCC =
    \frac{TP \cdot TN - FP \cdot FN}
    {\sqrt{(TP+FP)(TP+FN)(TN+FP)(TN+FN)}}.
\end{equation*}

\textbf{PR-AUC} measures classification performance by computing the area under the precision--recall curve, which characterizes the trade-off between precision and recall across different decision thresholds. In JIT-VP, high recall is desirable because missing a vulnerable commit can leave a vulnerability undetected, while high precision is important for reducing the number of false alarms that developers must investigate. PR-AUC summarizes this trade-off across decision thresholds and is therefore particularly appropriate for the highly imbalanced class distributions encountered in JIT-VP.

In this study, we evaluated 840 trained models and organized their results according to their corresponding VFC and VIC labeling strategies. We then performed pairwise comparisons between labeling strategies to assess the differences in their classification performance. For each comparison, we aggregated the paired results across the relevant models and projects, thereby accounting for variations in model architectures and dataset characteristics.

\subsubsection{\textbf{RQ2}} \textbf{Prioritization}

\textbf{Prioritization} aims to identify and prioritize commits that are most likely to be vulnerable for further developer analysis while explicitly accounting for the human effort required for inspection. To reflect this practical setting, we formulated JIT-VP prioritization as an \emph{effort-constrained ranking problem}. Given a predefined inspection budget, the objective is to identify as many vulnerable commits as possible while ensuring that the proportion of commits selected for developer inspection does not exceed the available budget. Specifically, for each model, we ranked commits in descending order of their predicted vulnerability probabilities. For a given effort budget, we then determined the prediction threshold that selects the largest possible set of commits without exceeding the specified budget and evaluated the model's performance under this constraint.

Unlike binary classification, however, evaluating an effort-constrained ranking task requires metrics that explicitly reflect the practical constraint of limited developer inspection effort. Conventional predictive metrics alone do not capture how effectively a model identifies vulnerable commits within a fixed inspection budget. Inspired by Zhou et al.~\cite{zhou2024comparisonstaticapplicationsecurity}, we therefore adopted effort-aware evaluation metrics, with adaptations to the JIT-VP setting, to assess the practical effectiveness of vulnerability ranking. Specifically, we considered two metrics: Vulnerability Detection Ratio (VDR), which measures the model's vulnerability detection performance, and Marked Commit Ratio (MCR), which quantifies the required developer inspection effort and serves as the constraint for selecting the prediction threshold.

\textbf{VDR} measures the proportion of vulnerable commits correctly identified by a model relative to the total number of vulnerable commits in the test set. It represents the security benefit of the model by quantifying how many truly vulnerable commits can be detected. Formally, VDR is defined as follows:
\begin{align*}
    \mathrm{VDR} &= \frac{N_{\text{vul\_detected}}}{N_{\text{vul\_total}}}.
\end{align*}

\textbf{MCR} measures the proportion of commits marked for inspection relative to the total number of commits in the test set. It represents the human inspection effort required to obtain the corresponding security benefit. For example, an MCR of 0.1 means that developers need to manually inspect 10\% of all commits. Formally, MCR is defined as follows:
\begin{align*}
    \mathrm{MCR} &= \frac{N_{\text{marked\_commit}}}{N_{\text{total\_commit}}}.
\end{align*}

Overall, a desirable JIT-VP model should achieve a high VDR while maintaining the MCR within the available inspection budget. A model with a high VDR may still be impractical if it also exhibits a high MCR, as this requires developers to inspect an excessive number of commits. Therefore, an effective JIT-VP approach must balance the trade-off between maximizing vulnerability detection and minimizing inspection effort.

In this study, we considered five inspection budgets: 5\%, 7.5\%, 10\%, 15\%, and 20\%. Combining these five budgets with the 840 trained models resulted in 4,200 analysis instances. For each instance, we selected the prediction threshold that achieved the highest VDR while ensuring that the resulting MCR did not exceed the specified inspection budget. We then performed pairwise comparisons of VDR between labeling strategies while preserving the pairing across the same model, project, and inspection budget. The resulting paired observations were subsequently aggregated across models, projects, and budgets to assess the effects of VFC and VIC labeling strategies on effort-aware vulnerability prioritization performance. Since MCR served only as a constraint for threshold selection, we report model performance solely in terms of VDR.

\subsubsection{\textbf{RQ3}} \textbf{Localization} 

\textbf{Localization} aims to pinpoint the potential root causes of vulnerabilities, thereby supporting developers in inspecting and patching vulnerable code. Depending on the granularity of the analysis, localization can be performed at the file, method, or line level. We formulated this JIT-VP task by evaluating the \emph{line-level localization} capability of JITFine, DeepJIT, and SimCom. Inspired by Fu et al.~\cite{fu2022linevul}, we assessed the models' localization capability by extracting the importance of individual input tokens and mapping these token-level importance scores to their corresponding lines of code.

To evaluate localization performance, we adopted Mean Reciprocal Rank \textbf{(MRR)} as the primary evaluation metric for the line-level explanation task. 
\textbf{MRR} measures how highly the first correctly identified vulnerable line is ranked by the explanation model in the given list of vulnerable commits according to their estimated relevance to the vulnerability. For a commit $i$, the reciprocal rank is defined as $1/r_i$, where $r_i$ denotes the rank of the highest-ranked line that belongs to the ground-truth vulnerable lines. The MRR over a set of $N$ vulnerable commits is then defined as:

\begin{equation*}
    \mathrm{MRR} = \frac{1}{N}\sum_{i=1}^{N}\frac{1}{r_i}.
\end{equation*}

In this study, we first applied the trained models of each architecture to 40 manually curated benchmark commits, comprising 28 Linux commits and 12 OpenSSL commits. For each commit predicted as vulnerable, we obtained the importance score of each input token and mapped these scores to their corresponding lines of code. We then aggregated the token-level scores within each line to derive a line-level vulnerability importance score. The lines of code were subsequently ranked in descending order of their importance scores, producing a ranked explanation of the model's prediction. Finally, we compared these ranked explanations against the manually annotated vulnerable lines described in Section~\ref{sec:benchmark} and reported the MRR metric to measure how effectively each model ranks the vulnerable lines near the top of its explanation.
\begin{table}[t]
\centering
\caption{Summary of the trained models in our experiments.}
\label{tab:configuration_space}
\renewcommand{\arraystretch}{1.3}
\setlength{\tabcolsep}{6pt}
\begin{tabular}{@{}lp{7cm}@{}}
\toprule
\textsc{Factor} & \textsc{Values} \\
\midrule
Model         & TLEL, LR, LAPredict, DeepJIT, SimCom, JITFine \\
% Budget        & 0.05, 0.075, 0.1, 0.15, 0.2 \\
% Dataset       & OpenSSL, Linux kernel \\
Sampling Seeds & 5 seeds \\
% Configuration & 28 configs of $\text{data}_{x\text{-}y}$ $x \in \{0,1,2,3\}$, $y \in \{0,1,2,3\}$ \\
Configurations & 16 Linux + 12 OpenSSL \\
\midrule
\textbf{Total} & \textbf{840} \\
\bottomrule
\end{tabular}
\end{table}

\subsection{Statistical Test and Effect Size}

To compare model performance across different experimental settings and datasets, we complemented the performance metrics with statistical significance testing and effect size analysis.
Since the performance measurements across configurations do not satisfy the normality assumption, we employed the Wilcoxon signed-rank test~\cite{woolson2007wilcoxon} to assess whether the observed differences between two labeling strategies are statistically significant. The Wilcoxon signed-rank test is a non-parametric paired test that evaluates whether the distribution of paired differences is systematically shifted from zero, making it suitable for comparing model performance under matched experimental settings. We considered a difference statistically significant when the resulting $p$-value is less than 0.05. For comparisons that do not reach statistical significance, we additionally reported the win rate and tie rate, calculated as $\text{Win}/\text{All}$ and $\text{Tie}/\text{All}$, respectively, to characterize the relative performance of the two strategies.

For RQ1, each paired observation corresponds to the same model, project, and random seed evaluated under two labeling configurations. For RQ2, paired observations are additionally matched by inspection budget, resulting in experimental units defined by model, project, random seed, and budget. For RQ3, each paired observation corresponds to the same model, project, and random seed under two labeling configurations. The Wilcoxon signed-rank tests therefore compare performance differences under matched experimental conditions.

In addition to statistical significance, we reported Cliff's delta~\cite{cliff1993dominance} ($\delta$) as an effect-size measure to quantify the magnitude and direction of performance differences between the two labeling strategies. Unlike significance tests, which assess whether an observed difference is unlikely to arise by chance, Cliff's delta measures the degree of dominance between two distributions and is robust to non-normal data. Following the guidelines proposed by Romano et al.~\cite{romano2006appropriate}, we interpreted the magnitude of $|\delta|$ as negligible ($<0.147$), small ($0.147$--$0.330$), medium ($0.330$--$0.474$), and large ($\geq0.474$). Positive $\delta$ values indicated that the automated labeling strategy tends to outperform the manual labeling strategy, whereas negative values indicate the opposite. We interpret statistical significance jointly with Cliff’s delta and the observed metric values, since a statistically significant difference does not necessarily imply a practically large improvement.

\subsection{Implementation Details}

For experiments, we constructed 28 data configurations, denoted as data$_{x-y}$, as described in Table~\ref{tab:design_data_stat}. Each configuration represented a distinct combination of VFC and VIC labeling strategies. We trained the studied JIT-VP models separately on each data configuration to systematically examine the impact of training data labels on model performance.

To conduct various experiments, we leveraged the implementations of models and label techniques from the original replication packages with their default parameters. All implemented models were trained using random undersampling (RUS) to address the severe class imbalance in realistic JIT-VP datasets. We adopted RUS following the recommendation of Nguyen et al.~\cite{nguyen2025toward}, who showed that it provides more effective performance than alternative sampling strategies under the substantial class imbalance encountered in realistic JIT-VP settings. To mitigate the influence of randomness in the sampling phase, we repeated the undersampling and training procedure using five different random seeds for each experimental configuration. 
% We subsequently aggregate the results across these five runs in our analysis.

We trained our studied models on an NVIDIA V100 GPU (32GB VRAM), 188GB RAM, and two 36-core Intel Xeon Gold 6140 CPUs (2.30GHz) running Ubuntu 24.04.3 LTS. Overall, this experimental procedure resulted in a total of 840 trained models across all data configurations, studied models, projects, and random seeds, as summarized in Table~\ref{tab:configuration_space}.

\section{Experiment Results}
\label{sec:evaluation}

In this section, we present the experimental results for each RQ, which respectively investigates the impact of different labeling techniques on the classification, prioritization, and localization tasks. Note that manually labeled VICs (VIC$_0$) are not available for OpenSSL. Therefore, comparisons involving VIC$_0$ in the following tables are restricted to the Linux kernel. Accordingly, the corresponding results for other VIC labeling techniques are reported using their Linux-only configurations in these comparisons.

\subsection{RQ1: Impact of Different Labeling Techniques on the Classification Task}

% \subsubsection{Evaluation Metrics}
% ============================================================
% VFC table (overall)
% ============================================================
\begin{table}[t]
\centering
\caption{Pairwise Wilcoxon signed-rank test results with Cliff's $\delta$ effect sizes for different VFC labeling strategies in the classification task using the MCC metric. The bold results show the higher mean performance configurations and statistically significant differences.}
\label{tab:rq1_vfc}
% \resizebox{\columnwidth}{!}{%
\begin{tabular}{ll cc c c l}
\toprule
\multicolumn{2}{c}{\textbf{Pair}} &
\multicolumn{2}{c}{\textbf{Mean} $MCC$} &
\textbf{$p$-value} &
\textbf{Cliff's $\delta$} &
\textbf{Magnitude} \\
% \cmidrule(lr){1-2}\cmidrule(lr){3-4}
\midrule
VFC$_{0}$ & VFC$_{1}$ & \textbf{0.0466} &         0.0434  & \textbf{0.003} & $+$0.065 & negligible \\
VFC$_{0}$ & VFC$_{2}$ & \textbf{0.0466} &         0.0364  & \textbf{0.004} & $+$0.242 & small \\
VFC$_{0}$ & VFC$_{3}$ & \textbf{0.0466} &         0.0421  & \textbf{0.001} & $+$0.118 & negligible \\
VFC$_{1}$ & VFC$_{2}$ & \textbf{0.0434} &         0.0364  & \textbf{0.046} & $+$0.172 & small \\
VFC$_{1}$ & VFC$_{3}$ & \textbf{0.0434} &         0.0421  & 0.059 & $+$0.043 & negligible \\
VFC$_{2}$ & VFC$_{3}$ &         0.0364  & \textbf{0.0421} & \textbf{0.048} & $-$0.116 & negligible \\
\bottomrule
\end{tabular}%
% }
\end{table}
% ============================================================
% VFC table (overall) - PR-AUC metric
% ============================================================
\begin{table}[t]
\centering
\caption{Pairwise Wilcoxon signed-rank test results with Cliff's $\delta$ effect sizes for different VFC labeling strategies in the classification task using the PR-AUC metric. The bold results show the higher mean performance configurations and statistically significant differences.}
\label{tab:rq1_vfc_prauc}
% \resizebox{\columnwidth}{!}{%
\begin{tabular}{ll cc c c l}
\toprule
\multicolumn{2}{c}{\textbf{Pair}} &
\multicolumn{2}{c}{\textbf{Mean} $PR\text{-}AUC$} &
\textbf{$p$-value} &
\textbf{Cliff's $\delta$} &
\textbf{Magnitude} \\
% \cmidrule(lr){1-2}\cmidrule(lr){3-4}
\midrule
VFC$_{0}$ & VFC$_{1}$ & \textbf{0.0203} &         0.0179  & 0.112 & $+$0.007 & negligible \\
VFC$_{0}$ & VFC$_{2}$ & \textbf{0.0203} &         0.0177  & 0.181 & $+$0.044 & negligible \\
VFC$_{0}$ & VFC$_{3}$ & \textbf{0.0203} &         0.0177  & \textbf{0.007} & $+$0.037 & negligible \\
VFC$_{1}$ & VFC$_{2}$ & \textbf{0.0179} &         0.0177  & 0.477 & $+$0.048 & negligible \\
VFC$_{1}$ & VFC$_{3}$ & \textbf{0.0179} &         0.0177  & 0.280 & $+$0.026 & negligible \\
VFC$_{2}$ & VFC$_{3}$ &         0.0177  & 0.0177 & 0.494 & $-$0.061 & negligible \\
\bottomrule
\end{tabular}%
% }
\end{table}

Table~\ref{tab:rq1_vfc} and Table~\ref{tab:rq1_vfc_prauc} present the pairwise comparisons of VFC labeling strategies using MCC and PR-AUC, respectively, while the corresponding results for VIC labeling strategies are shown in Table~\ref{tab:rq1_vic} and Table~\ref{tab:rq1_vic_prauc}. 
% Overall, the two evaluation metrics reveal different levels of sensitivity to labeling strategies. MCC shows more pronounced differences, particularly among VFC strategies, whereas PR-AUC indicates that most labeling strategies achieve comparable classification performance.

% ============================================================
% VIC table (overall)
% ============================================================
\begin{table}[t]
\centering
\caption{Pairwise Wilcoxon signed-rank test results with Cliff's $\delta$ effect sizes for different VIC labeling strategies in the classification task using MCC metric. The bold results show the higher mean performance configurations and statistically significant differences. The $*$ symbol denotes results available only for the Linux kernel.}
\label{tab:rq1_vic}
% \resizebox{\columnwidth}{!}{%
\begin{tabular}{ll cc c c l}
\toprule
\multicolumn{2}{c}{\textbf{Pair}} &
\multicolumn{2}{c}{\textbf{Mean} $MCC$} &
\textbf{$p$-value} &
\textbf{Cliff's $\delta$} &
\textbf{Magnitude} \\
% \cmidrule(lr){1-2}\cmidrule(lr){3-4}
\midrule
VIC$^*_{0}$ & VIC$^*_{1}$ &          0.0247  & \textbf{0.0255} & 0.459 & $+$0.028 & negligible \\
VIC$^*_{0}$ & VIC$^*_{2}$ &          0.0247  & \textbf{0.0265} & 0.194 & $+$0.104 & negligible \\
VIC$^*_{0}$ & VIC$^*_{3}$ & \textbf{0.0247}  &         0.0232  & 0.156 & $+$0.104 & negligible \\
VIC$_{1}$ & VIC$_{2}$ & 0.0457  &         0.0457  & 0.419 & $+$0.004 & negligible \\
VIC$_{1}$ & VIC$_{3}$ & \textbf{0.0457}  &         0.0437  & \textbf{0.018} & $+$0.075 & negligible \\
VIC$_{2}$ & VIC$_{3}$ & \textbf{0.0457}  &         0.0437  & \textbf{0.040} & $+$0.060 & negligible \\
\bottomrule
\end{tabular}%
% }
\end{table}
% ============================================================
% VIC table (overall) - PR-AUC metric
% ============================================================
\begin{table}[t]
\centering
\caption{Pairwise Wilcoxon signed-rank test results with Cliff's $\delta$ effect sizes for different VIC labeling strategies in the classification task using PR-AUC metric. The bold results show the higher mean performance configurations and statistically significant differences. The $*$ symbol denotes results available only for the Linux kernel. }
\label{tab:rq1_vic_prauc}
% \resizebox{\columnwidth}{!}{%
\begin{tabular}{ll cc c c l}
\toprule
\multicolumn{2}{c}{\textbf{Pair}} &
\multicolumn{2}{c}{\textbf{Mean} $PR\text{-}AUC$} &
\textbf{$p$-value} &
\textbf{Cliff's $\delta$} &
\textbf{Magnitude} \\
% \cmidrule(lr){1-2}\cmidrule(lr){3-4}
\midrule
VIC$^*_{0}$ & VIC$^*_{1}$ &         0.0038  & \textbf{0.0055} & 0.057 & $+$0.174 & small \\
VIC$^*_{0}$ & VIC$^*_{2}$ &         0.0038  & \textbf{0.0048} & \textbf{0.013} & $+$0.236 & small \\
VIC$^*_{0}$ & VIC$^*_{3}$ &         0.0038  & \textbf{0.0043} & 0.066 & $+$0.132 & negligible \\
VIC$_{1}$ & VIC$_{2}$ &         0.0209  & \textbf{0.0216} & 0.221 & $+$0.017 & negligible \\
VIC$_{1}$ & VIC$_{3}$ & \textbf{0.0209} &         0.0200  & 0.220 & $+$0.028 & negligible \\
VIC$_{2}$ & VIC$_{3}$ & \textbf{0.0216} &         0.0200  & 0.058 & $+$0.039 & negligible \\
\bottomrule
\end{tabular}%
% }
\end{table}

\paragraph{\textbf{VFC labeling strategies}}
In terms of MCC, manual VFC labeling (VFC$_0$) provides the strongest classification performance, whereas the quantity-first strategy (VFC$_2$) consistently performs the worst. Specifically, VFC$_0$ achieves the highest mean MCC of 0.0466, followed by VFC$_1$, VFC$_3$, and VFC$_2$. Moreover, VFC$_0$ significantly outperforms all three alternative strategies. Meanwhile, there is little difference between VFC$_1$ and VFC$_3$, but both strategies significantly outperform VFC$_2$, with $p=0.046$ and $p=0.048$, respectively. These results suggest that better quality VFC labeling is beneficial for classification performance as measured by MCC.
% , whereas prioritizing the quantity of identified VFCs leads to the weakest performance.

However, the PR-AUC results present a different picture. VFC$_0$ achieves the highest mean PR-AUC (0.0203), followed by VFC$_1$, while VFC$_2$ and VFC$_3$ achieve comparable performance. Despite its higher mean performance, VFC$_0$ significantly outperforms only VFC$_3$ and does not achieve a statistically significant improvement over VFC$_1$ or VFC$_2$. Furthermore, none of the remaining pairwise comparisons among the automated strategies is
statistically significant.

The discrepancy between the MCC and PR-AUC results may be explained by the different aspects of classification performance captured by the two metrics. MCC incorporates all four outcomes of binary classification, including true negatives, whereas PR-AUC focuses exclusively on the trade-off between precision and recall for the positive class. In the practical setting, where non-vulnerable commits substantially outnumber vulnerable ones, improved VFC labels may primarily help models better distinguish benign commits from vulnerable ones, leading to improvements in MCC. While the limited differences in PR-AUC suggest that improving VFC labels does not necessarily translate into a corresponding improvement in the ability to identify the relatively scarce vulnerable commits. In other words, better VFC labeling appears to have a stronger effect on overall classification discrimination than on learning patterns associated with the positive class.

% \begin{summarybox}[title=\textbf{Answer to RQ1 - 1}]
% High-quality VFC labels significantly improve model performance in terms of MCC, but provide limited improvement in PR-AUC. This discrepancy may stem from the severe class imbalance in realistic JIT-VP datasets, where improved VFC labels can help models better distinguish the dominant non-vulnerable class while providing limited benefit in learning patterns associated with the scarce vulnerable class.
% \end{summarybox}

% Add this to your preamble if not already present:
% \usepackage{threeparttable}
% \usepackage{multirow}
% \usepackage{booktabs}
\begin{table*}[t]
\centering
\caption{Pairwise Wilcoxon signed-rank test results with Cliff's $\delta$ effect sizes of recommended combination, data$_{0-1}$ (VFC$_0$, VIC$_1$), against all other dataset configurations data$_{x-y}$ in the classification task using MCC metric. The bold results show the highest mean performance configurations and statistically significant differences. The $*$ symbol denotes results available only for the Linux kernel.}
\label{tab:rq1_best_combo_mcc}
\begin{threeparttable}
\begin{tabular}{@{}l r r r l r r@{}}
\toprule
\multirow{2}{*}{\textbf{Baselines}} &
  \multirow{2}{*}{\textbf{Mean}} &
  \multirow{2}{*}{\textbf{\textit{p}-value}} &
  \multicolumn{2}{c}{\textbf{Cliff's $\delta$}} &
  \multirow{2}{*}{\textbf{Win\,\%}} &
  \multirow{2}{*}{\textbf{Tie\,\%}} \\
\cmidrule(lr){4-5}
  & & & {\textbf{Value}} & {\textbf{Magnitude}} & & \\
\midrule
data$^*_{0-1}$ & \textbf{0.0302} & -- & -- & -- & -- & -- \\
data$_{0-1}$   & 0.0513          & -- & -- & -- & -- & -- \\
\midrule
%--- VFC-0 group ---
\multicolumn{7}{@{}l}{} \\[-6pt]
data$^*_{0-0}$  & 0.0243          & 0.0625          & 0.3056 & Small      & 66.67 & 16.67 \\
data$^*_{1-0}$  & 0.0237 & 0.0938          & 0.3611 & Medium     & 66.67 & 16.67 \\
data$^*_{2-0}$  & 0.0243 & 0.1563          & 0.2500 & Small  & 66.67 & 16.67 \\
data$^*_{3-0}$  & 0.0265 & \textbf{0.0313} & 0.3056 & Small & 83.33 & 16.67 \\

%--- VFC-1 group ---
\multicolumn{7}{@{}l}{} \\[-6pt]

data$_{1-1}$  & 0.0507 & 0.3188          & 0.0069 & Negligible & 58.33 & 8.33  \\
data$_{2-1}$  & 0.0349 & \textbf{0.0137} & 0.3611 & Medium & 75.0  & 16.67 \\
data$_{3-1}$  & 0.0459 & \textbf{0.0322} & 0.1528 & Small & 66.67 & 16.67 \\

%--- VFC-2 group ---
\multicolumn{7}{@{}l}{} \\[-6pt]

data$_{0-2}$  & \textbf{0.0514} & 0.4229          & 0.0000 & Negligible & 41.67 & 16.67 \\
data$_{1-2}$  & 0.0446 & \textbf{0.0420} & 0.1389 & Negligible & 58.33 & 16.67 \\
data$_{2-2}$  & 0.0415 & 0.1611          & 0.2361 & Small  & 50.0  & 16.67 \\
data$_{3-2}$  & 0.0452 & \textbf{0.0322} & 0.1806 & Small & 66.67 & 16.67 \\

%--- VFC-3 group ---
\multicolumn{7}{@{}l}{} \\[-6pt]
data$_{0-3}$  & 0.0483          & \textbf{0.0186} & 0.1528 & Small      & 75.0  & 16.67 \\
data$_{1-3}$  & 0.0449 & \textbf{0.0020} & 0.2083 & Small      & 75.0  & 16.67 \\
data$_{2-3}$  & 0.0387 & \textbf{0.0322} & 0.2917 & Small  & 66.67 & 16.67 \\
data$_{3-3}$  & 0.0430 & \textbf{0.0010} & 0.2361 & Small & 83.33 & 16.67 \\
\bottomrule
\end{tabular}
\begin{tablenotes}
\footnotesize
\item data$^*_{0-1}$ reports the Linux-only mean for comparisons with Linux-only results \\ data$_{0-1}$ reports the overall mean for other comparisons.
\end{tablenotes}
\end{threeparttable}
\end{table*}
% Add this to your preamble if not already present:
% \usepackage{threeparttable}
% \usepackage{multirow}
% \usepackage{booktabs}
\begin{table*}[t]
\centering
\caption{Pairwise Wilcoxon signed-rank test results with Cliff's $\delta$ effect sizes of recommended combination, data$_{0-2}$ (VFC$_0$, VIC$_2$) against all other dataset configurations data$_{x-y}$ in the classification task using PR-AUC metric. The bold results show the highest mean performance configurations and statistically significant differences. The $*$ symbol denotes results available only for the Linux kernel.}
\label{tab:rq1_best_combo_prauc_02}
\begin{threeparttable}
\begin{tabular}{@{}l r r r l r r@{}}
\toprule
\multirow{2}{*}{\textbf{Baselines}} &
  \multirow{2}{*}{\textbf{Mean}} &
  \multirow{2}{*}{\textbf{\textit{p}-value}} &
  \multicolumn{2}{c}{\textbf{Cliff's $\delta$}} &
  \multirow{2}{*}{\textbf{Win\,\%}} &
  \multirow{2}{*}{\textbf{Tie\,\%}} \\
\cmidrule(lr){4-5}
  & & & {\textbf{Value}} & {\textbf{Magnitude}} & & \\
\midrule
data$^*_{0-2}$ & 0.0042 & -- & -- & -- & -- & -- \\
data$_{0-2}$   & \textbf{0.0235} & -- & -- & -- & -- & -- \\
\midrule
%--- VFC-0 group ---
\multicolumn{7}{@{}l}{} \\[-6pt]
data$^*_{0-0}$  & 0.0033          & 0.1562 & 0.4167 & Medium     & 66.67 & 16.67 \\
data$^*_{1-0}$  & 0.0033 & 0.1562 & 0.3611 & Medium     & 50.0  & 16.67 \\
data$^*_{2-0}$  & \textbf{0.0046} & 0.1562 & 0.2500 & Small      & 16.67 & 16.67 \\
data$^*_{3-0}$  & 0.0039 & 0.2188          & 0.2500 & Small      & 66.67 & 16.67 \\

%--- VFC-1 group ---
\multicolumn{7}{@{}l}{} \\[-6pt]
data$_{0-1}$  & 0.0224          & 0.3188 & 0.0208 & Negligible & 41.67 & 8.33  \\
data$_{1-1}$  & 0.0218 & 0.0737 & 0.0625 & Negligible & 66.67 & 8.33  \\
data$_{2-1}$  & 0.0187 & 0.1201 & 0.0486 & Negligible & 58.33 & 8.33  \\
data$_{3-1}$  & 0.0209 & \textbf{0.0098} & 0.0972 & Negligible & 66.67 & 16.67 \\

%--- VFC-2 group ---
\multicolumn{7}{@{}l}{} \\[-6pt]

data$_{1-2}$  & 0.0195 & 0.3848 & 0.0278 & Negligible & 33.33 & 16.67 \\
data$_{2-2}$  & 0.0233 & 0.3501 & 0.0208 & Negligible & 41.67 & 8.33  \\
data$_{3-2}$  & 0.0202 & 0.0654          & 0.0556 & Negligible & 58.33 & 16.67 \\

%--- VFC-3 group ---
\multicolumn{7}{@{}l}{} \\[-6pt]
data$_{0-3}$  & 0.0234 & 0.1392 & 0.0764 & Negligible & 66.67 & 8.33  \\
data$_{1-3}$  & 0.0198 & 0.0654 & 0.0417 & Negligible & 58.33 & 16.67 \\
data$_{2-3}$  & 0.0179 & 0.0874 & 0.0486 & Negligible & 58.33 & 8.33  \\
data$_{3-3}$  & 0.0191 & \textbf{0.0034} & 0.1319 & Negligible & 75.0  & 8.33  \\
\bottomrule
\end{tabular}
\begin{tablenotes}
\footnotesize
\item data$^*_{0-2}$ reports the Linux-only mean for comparisons with Linux-only results \\ data$_{0-2}$ reports the overall mean for other comparisons.
\end{tablenotes}
\end{threeparttable}
\end{table*}

\paragraph{\textbf{VIC labeling strategies.}}
The MCC results for VIC labeling reveal relatively limited differences among the strategies. For the comparisons, VIC$_0$ is slightly outperformed by VIC$_1$ and VIC$_2$, but neither difference is statistically significant with $p=0.459$ and $p=0.194$, respectively. VIC$_3$ has a lower mean MCC than VIC$_0$, but this difference is also not statistically significant.

The same picture can be observed in the PR-AUC results. VIC$_0$ achieves the lowest mean PR-AUC of 0.0038, while VIC$_1$ and VIC$_2$ achieve higher values of 0.0055 and 0.0048, respectively. 
None of the other pairwise comparisons are statistically significant except for the difference between VIC$_2$ and VIC$_0$ with $p=0.013$.

Taken together, the results indicate that VIC labeling strategies have a relatively modest effect on downstream classification performance. It suggests that improving VIC label quality alone may be insufficient to substantially improve JIT-VP classification performance, motivating further investigation into model architecture, learning strategies, and the ability of existing models to capture vulnerability-related patterns.

\paragraph{\textbf{Recommended combination}}

The best performing configuration in terms of MCC is data$_{0-1}$, as reported in Table~\ref{tab:rq1_best_combo_mcc}. It achieves a mean MCC of 0.0513 and significantly outperforms nine other configurations. Although data$_{0-2}$ achieves a nearly identical mean MCC (0.0514), its difference from data$_{0-1}$ is not significant ($p=0.4229$) and it does not achieve the same number of significant wins. Notably, data$_{1-1}$, which uses fully automated VFC and VIC labeling, also achieves a competitive mean MCC of 0.0507 without a significant difference from data$_{0-1}$ ($p=0.3188$). This result suggests that fully automated labeling can be a viable and scalable alternative for constructing JIT-VP datasets, particularly for large or poorly documented projects.

Table~\ref{tab:rq1_best_combo_prauc_02} reveals a different pattern for PR-AUC. As discussed above, the models struggle to consistently identify positive instances across decision thresholds, resulting in very low PR-AUC values that are strongly affected by the severe class imbalance. In particular, data$^*_{0-2}$ achieves a substantially lower PR-AUC than data$_{0-2}$ since class imbalance is more extreme in the Linux dataset. Although data$_{0-2}$ achieves the highest mean PR-AUC (0.0235), it significantly outperforms only data$_{3-1}$ ($p=0.0098$) and data$_{3-3}$ ($p=0.0034$). The absence of significant differences against most configurations makes it difficult to identify a definitive best configuration based on PR-AUC. This result further highlights the difficulty of learning discriminative patterns of vulnerable commits under extreme class imbalance in JIT-VP, where improvements in overall classification performance do not necessarily translate into better identification of the minority vulnerable class.

Overall, the results favor manual VFC labeling combined with vulnerability-specific VIC labeling (data$_{0-1}$) as the most reliable configuration for the classification task, particularly in terms of MCC. 

\begin{summarybox}[title=\textbf{Answer to RQ1}]
VFC label quality has a stronger effect on MCC than on PR-AUC, suggesting that improved labels enhance overall classification but provide limited benefit for learning patterns of the scarce vulnerable class under severe class imbalance.
VIC labeling strategies have a comparatively modest impact on both metrics, indicating that label refinement alone may be insufficient to substantially improve JIT-VP performance.
Overall, data$_{0-1}$ (manual VFC and vulnerability-specific VIC) provides the most reliable configuration, particularly for MCC, while fully automated labeling remains a promising scalable alternative.
\end{summarybox}

\subsection{RQ2: Impact of Different Labeling Techniques on the Prioritization Task}

Tables~\ref{tab:rq2_vfc} and~\ref{tab:rq2_vic} present the results of pairwise comparison of different labeling strategies in the ranking task with VDR metric. 

\begin{table}[t]
\centering
\caption{Pairwise Wilcoxon signed-rank test results with Cliff's $\delta$ effect sizes for different VFC labeling strategies in the prioritization task using VDR metric. The bold results show the higher mean performance configurations and statistically significant differences.}
\label{tab:rq2_vfc}
% \resizebox{\columnwidth}{!}{%
\begin{tabular}{ll cc c c l}
\toprule
\multicolumn{2}{c}{\textbf{Pair}} &
\multicolumn{2}{c}{\textbf{Mean} $VDR$} &
\textbf{$p$-value} &
\textbf{Cliff's $\delta$} &
\textbf{Magnitude} \\
% \cmidrule(lr){1-2}\cmidrule(lr){3-4}
% \textbf{$A$} & \textbf{$B$} & \textbf{$A$} & \textbf{$B$} & & & \\
\midrule
VFC$_{0}$ & VFC$_{1}$ & \textbf{0.382} &          0.378 & 0.068 & $+$0.018 & negligible \\
VFC$_{0}$ & VFC$_{2}$ & \textbf{0.382} &          0.342 & \textbf{3.32$e-$7}    & $+$0.155 & small \\
VFC$_{0}$ & VFC$_{3}$ & \textbf{0.382}  &         0.367 & \textbf{1.77$e-$6} & $+$0.060 & negligible \\
VFC$_{1}$ & VFC$_{2}$ & \textbf{0.378} &          0.342 & \textbf{9.67$e-$6} & $+$0.132 & negligible \\
VFC$_{1}$ & VFC$_{3}$ & \textbf{0.378} &          0.367 & \textbf{5.01$e-$5}    & $+$0.041 & negligible \\
VFC$_{2}$ & VFC$_{3}$ &         0.342  & \textbf{0.367} & \textbf{0.0001} & $+$0.092 & negligible \\
\bottomrule
\end{tabular}%
% }
\end{table}

\begin{table}[t]
\centering
\caption{Pairwise Wilcoxon signed-rank test results with Cliff's $\delta$ effect sizes for different VIC labeling strategies in the prioritization task using VDR metric. The bold results show the higher mean performance configurations and statistically significant differences. The $*$ symbol denotes results available only for the Linux kernel.}
\label{tab:rq2_vic}
% \resizebox{\columnwidth}{!}{%
\begin{tabular}{ll cc c c l}
\toprule
\multicolumn{2}{c}{\textbf{Pair}} &
\multicolumn{2}{c}{\textbf{Mean} $VDR$} &
\textbf{$p$-value} &
\textbf{Cliff's $\delta$} &
\textbf{Magnitude} \\
% \cmidrule(lr){1-2}\cmidrule(lr){3-4}
% \textbf{$A$} & \textbf{$B$} & \textbf{$A$} & \textbf{$B$} & & & \\
\midrule
VIC$^*_{0}$ & VIC$^*_{1}$ & 0.372 & 0.372 & 0.260 & $-$0.011 & negligible \\
VIC$^*_{0}$ & VIC$^*_{2}$ & 0.372 & 0.372 & 0.167 & $-$0.003 & negligible \\
VIC$^*_{0}$ & VIC$^*_{3}$ & \textbf{0.372} & 0.363 & \textbf{0.041} & $+$0.029 & negligible \\
VIC$_{1}$ & VIC$_{2}$ & \textbf{0.369} & 0.368 & 0.420 & $+$0.002 & negligible \\
VIC$_{1}$ & VIC$_{3}$ & \textbf{0.369} & 0.364 & \textbf{0.011}   & $+$0.021 & negligible \\
VIC$_{2}$ & VIC$_{3}$ & \textbf{0.368} & 0.363 & \textbf{0.032} & $+$0.020 & negligible \\
\bottomrule
\end{tabular}%
% }
\end{table}

% Overall, the results indicate that the choice of VFC labeling strategy has a statistically significant effect on ranking performance for most pairwise comparisons, but the magnitude of these differences is generally small.

\paragraph{\textbf{VFC labeling strategies.}}
VFC labeling strategies have a significant impact on the performance of JIT-VP models in the effort-aware ranking task. In terms of mean VDR, the strategies follow the ordering VFC$_0$, VFC$_1$, VFC$_3$ and VFC$_2$, with manual labeling (VFC$_0$) achieving the highest VDR and the quantity-first strategy (VFC$_2$) achieving the lowest. All pairwise comparisons are statistically significant except for the comparison between VFC$_0$ and VFC$_1$ ($p=0.068$). This result suggests that VFC$_0$ is the most favourable strategy for effort-aware ranking, while VFC$_1$ provides statistically comparable performance to manual labeling. In contrast, the quantity-first strategy does not translate into better ranking performance. These findings indicate that, when constructing JIT-VP training data for ranking, researchers should prioritize VFC label accuracy over the quantity of identified vulnerable commits.

\paragraph{\textbf{VIC labeling strategies.}}
Unlike the VFC results, VIC labeling strategies exhibit little difference in ranking performance. VIC$_0$, VIC$_1$, and VIC$_2$ achieve comparable VDR scores, with none of their pairwise differences being statistically significant. In contrast, all three strategies significantly outperform VIC$_3$, the LLM-assisted strategy. Although LLM-assisted labeling provides a more sophisticated alternative to traditional approaches, its higher complexity does not translate into improved effort-aware ranking performance. These findings suggest that, despite their limitations as heuristic approaches, including their reliance on tracing back from VFCs and their inability to identify VICs for certain types of VFCs, B-SZZ and V-SZZ approaches provide a sufficiently effective and scalable solution for constructing VIC labels for the prioritization task.

% Add this to your preamble if not already present:
% \usepackage{threeparttable}
% \usepackage{multirow}
% \usepackage{booktabs}
\begin{table*}[t]
\centering
\caption{Pairwise Wilcoxon signed-rank test results with Cliff's $\delta$ effect sizes of recommended combination, data$_{0-1}$ (VFC$_0$, VIC$_1$) against all other dataset configurations data$_{x-y}$ in the prioritization task using VDR metric. The bold results show the highest mean performance configurations and statistically significant differences. The $*$ symbol denotes results available only for the Linux kernel.}
\label{tab:rq1_best_combo_vdr}
\begin{threeparttable}
\begin{tabular}{@{}l r r r l r r@{}}
\toprule
\multirow{2}{*}{\textbf{Baselines}} &
  \multirow{2}{*}{\textbf{Mean}} &
  \multirow{2}{*}{\textbf{\textit{p}-value}} &
  \multicolumn{2}{c}{\textbf{Cliff's $\delta$}} &
  \multirow{2}{*}{\textbf{Win\,\%}} &
  \multirow{2}{*}{\textbf{Tie\,\%}} \\
\cmidrule(lr){4-5}
  & & & {\textbf{Value}} & {\textbf{Magnitude}} & & \\
\midrule
data$^*_{0-1}$ & \textbf{0.3977} & -- & -- & -- & -- & -- \\
data$_{0-1}$   & \textbf{0.3917} & -- & -- & -- & -- & -- \\
\midrule
%--- VFC-0 group ---
\multicolumn{7}{@{}l}{} \\[-6pt]
data$^*_{0-0}$  & 0.3607          & \textbf{0.0001} & 0.1367 & Negligible & 60.0  & 16.67 \\
data$^*_{1-0}$  & 0.3684 & \textbf{0.0197}   & 0.1122 & Negligible & 53.33 & 16.67 \\
data$^*_{2-0}$  & 0.3771 & 0.0567            & 0.0944 & Negligible & 53.33 & 16.67 \\
data$^*_{3-0}$  & 0.3817 & \textbf{0.0226}   & 0.0644 & Negligible & 56.67 & 16.67 \\

%--- VFC-1 group ---
\multicolumn{7}{@{}l}{} \\[-6pt]

data$_{1-1}$  & 0.3884 & 0.0657            & 0.0125 & Negligible & 50.0  & 18.33 \\
data$_{2-1}$  & 0.3272 & \textbf{$<$0.0001} & 0.2458 & Small      & 61.67 & 18.33 \\
data$_{3-1}$  & 0.3689 & \textbf{0.0002}   & 0.0903 & Negligible & 60.0  & 20.0  \\

%--- VFC-2 group ---
\multicolumn{7}{@{}l}{} \\[-6pt]
data$_{0-2}$  & 0.3886 & 0.3706          & 0.0178 & Negligible & 40.0  & 20.0  \\
data$_{1-2}$  & 0.3822 & 0.0784            & 0.0381 & Negligible & 45.0  & 20.0  \\
data$_{2-2}$  & 0.3374 & \textbf{0.0007}   & 0.1958 & Small      & 55.0  & 20.0  \\
data$_{3-2}$  & 0.3626 & \textbf{$<$0.0001} & 0.1094 & Negligible & 60.0  & 20.0  \\

%--- VFC-3 group ---
\multicolumn{7}{@{}l}{} \\[-6pt]
data$_{0-3}$  & 0.3774          & \textbf{0.0016} & 0.0483 & Negligible & 48.33 & 18.33 \\
data$_{1-3}$  & 0.3696 & \textbf{$<$0.0001} & 0.0961 & Negligible & 61.67 & 18.33 \\
data$_{2-3}$  & 0.3446 & \textbf{0.0001}   & 0.1825 & Small      & 61.67 & 18.33 \\
data$_{3-3}$  & 0.3636 & \textbf{$<$0.0001} & 0.1067 & Negligible & 71.67 & 18.33 \\
\bottomrule
\end{tabular}
\begin{tablenotes}
\footnotesize
\item data$^*_{0-1}$ reports the Linux-only mean for comparisons with Linux-only results \\ data$_{0-1}$ reports the overall mean for other comparisons.
\end{tablenotes}
\end{threeparttable}
\end{table*}

\paragraph{\textbf{Recommended combination}}
Table~\ref{tab:rq1_best_combo_vdr} compares the dataset configurations in terms of VDR for the vulnerability prioritization task. Overall, data$_{0-1}$ is the best-performing configuration, achieving a mean VDR of 0.3917 and significantly outperforming most alternative configurations. The Linux-only result, data$^*_{0-1}$, is slightly higher (0.3977), indicating that the observed advantage of data$_{0-1}$ is not driven solely by the inclusion of OpenSSL.

Notably, several configurations exhibit statistically significant degradation compared with data$_{0-1}$. The strongest statistical evidence is observed for data$_{2-1}$ ($p<0.0001$), data$_{3-2}$ ($p<0.0001$), data$_{1-3}$ ($p<0.0001$), and data$_{3-3}$ ($p<0.0001$), followed by data$_{2-3}$ ($p=0.0001$) and data$_{0-0}$ ($p=0.0001$). The degradation is mainly observed in configurations with high label quantity but potentially noisier labels (VFC$_2$), or configurations with high priority on label accuracy but limited coverage (VIC$_0$), showing negative effects on vulnerability prioritization. This finding highlights the importance of balancing label quality and coverage when constructing JIT-VP datasets, especially for the prioritization task.

Overall, data$_{0-1}$ provides the most reliable labeling configuration for the prioritization task. The consistent performance degradation observed under configurations with either noisier or more limited labels further emphasizes the importance of maintaining a balance between label quality and quantity when constructing JIT-VP datasets.

\begin{summarybox}[title=\textbf{Answer to RQ2}]
VFC label quality has a stronger impact on effort-aware vulnerability ranking than label quantity, with noisy VFC labels consistently degrading performance. Thus, accurate VFC labeling should be prioritized when constructing ranking datasets.
SZZ-based VIC labeling achieves performance comparable to manual labeling and generally outperforms LLM-assisted labeling, making SZZ a practical and scalable approach for VIC construction.
Overall, data$_{0-1}$ provides the most reliable configuration for vulnerability prioritization, highlighting the importance of balancing label quality and coverage.
\end{summarybox}

% \input{plots/rq2_best_combo}

% In contrast, the LLM-assisted strategy VIC$_3$ achieves a lower mean VDR of 0.363, compared with 0.372 for each of the other three strategies. This difference is statistically significant when compared with VIC$_0$ ($p=0.041$, $\delta=0.029$), VIC$_1$ ($p=0.011$, $\delta=0.021$), and VIC$_2$ ($p=0.032$, $\delta=0.020$). Nevertheless, all three comparisons exhibit negligible effect sizes, indicating that the practical magnitude of the difference is small despite statistical significance.

% Overall, the results indicate that VIC labeling has a relatively limited effect on effort-aware ranking performance. Manual labeling (VIC$_0$), V-SZZ (VIC$_1$), and B-SZZ (VIC$_2$) produce virtually identical results, while the LLM-assisted approach (VIC$_3$) consistently underperforms the other strategies. Thus, unlike the VFC results, where the precision-first strategy provides a promising alternative to manual labeling, the traditional automated SZZ strategies already provide ranking performance comparable to manual VIC labeling without requiring manual annotation.

% ============================================================
% VFC table (overall) - MRR metric
% ============================================================
\begin{table}[t]
\centering
\caption{Pairwise Wilcoxon signed-rank test results with Cliff's $\delta$ effect sizes for different VFC labeling strategies in the localization task using MRR metric. The bold results show the higher mean performance configurations and statistically significant differences.}
\label{tab:rq1_vfc_mrr}
% \resizebox{\columnwidth}{!}{%
\begin{tabular}{ll cc c c l}
\toprule
\multicolumn{2}{c}{\textbf{Pair}} &
\multicolumn{2}{c}{\textbf{Mean} $MRR$} &
\textbf{$p$-value} &
\textbf{Cliff's $\delta$} &
\textbf{Magnitude} \\
% \cmidrule(lr){1-2}\cmidrule(lr){3-4}
\midrule
VFC$_{0}$ & VFC$_{1}$ & \textbf{0.3932} &         0.3841  & \textbf{0.023} & $+$0.077 & negligible \\
VFC$_{0}$ & VFC$_{2}$ & \textbf{0.3932} &         0.3825  & \textbf{8.06$e-$5} & $-$0.161 & small \\
VFC$_{0}$ & VFC$_{3}$ & \textbf{0.3932} &         0.3856  & \textbf{0.012} & $+$0.111 & negligible \\
VFC$_{1}$ & VFC$_{2}$ & \textbf{0.3841} &         0.3825  & \textbf{0.015} & $+$0.066 & negligible \\
VFC$_{1}$ & VFC$_{3}$ &         0.3841  & \textbf{0.3856} & 0.354 & $+$0.043 & negligible \\
VFC$_{2}$ & VFC$_{3}$ &         0.3825  & \textbf{0.3856} & 0.152 & $+$0.016 & negligible \\
\bottomrule
\end{tabular}%
% }
\end{table}
% ============================================================
% VIC table (overall) - MRR metric
% ============================================================
\begin{table}[t]
\centering
\caption{Pairwise Wilcoxon signed-rank test results with Cliff's $\delta$ effect sizes for different VIC labeling strategies in the localization task using MRR metric. The bold results show the higher mean performance configurations and statistically significant differences. The $*$ symbol denotes results available only for the Linux kernel.}
\label{tab:rq1_vic_mrr}
% \resizebox{\columnwidth}{!}{%
\begin{tabular}{ll cc c c l}
\toprule
\multicolumn{2}{c}{\textbf{Pair}} &
\multicolumn{2}{c}{\textbf{Mean} $MRR$} &
\textbf{$p$-value} &
\textbf{Cliff's $\delta$} &
\textbf{Magnitude} \\
% \cmidrule(lr){1-2}\cmidrule(lr){3-4}
\midrule
VIC$^*_{0}$ & VIC$^*_{1}$ & \textbf{0.3790} &         0.3778  & 0.259 & $+$0.097 & negligible \\
VIC$^*_{0}$ & VIC$^*_{2}$ &         0.3790  & \textbf{0.3800} & 0.396 & $+$0.062 & negligible \\
VIC$^*_{0}$ & VIC$^*_{3}$ &         0.3790  & \textbf{0.3820} & 0.102 & $+$0.167 & small \\
VIC$_{1}$ & VIC$_{2}$ & \textbf{0.3861} &         0.3836  & 0.439 & $+$0.017 & negligible \\
VIC$_{1}$ & VIC$_{3}$ &         0.3861  & \textbf{0.3929} & \textbf{0.004} & $+$0.090 & negligible \\
VIC$_{2}$ & VIC$_{3}$ &         0.3836  & \textbf{0.3929} & \textbf{0.001} & $+$0.080 & negligible \\
\bottomrule
\end{tabular}%
% }
\end{table}

\subsection{RQ3: Impact of Different Labeling Techniques on the Localization Task}

We evaluate the impact of VFC and VIC labeling strategies on the quality of line-level vulnerability explanations using Mean Reciprocal Rank (MRR). Table~\ref{tab:rq1_vfc_mrr} and Table~\ref{tab:rq1_vic_mrr} present the pairwise comparisons for VFC and VIC labeling strategies, respectively.

\paragraph{\textbf{VFC labeling strategies}}
As shown in Table~\ref{tab:rq1_vfc_mrr}, VFC$_0$ achieves the highest overall mean MRR of 0.3932, followed by VFC$_3$, VFC$_1$, and VFC$_2$. Manual labeling (VFC$_0$) significantly outperforms all three alternative strategies. Among the automated strategies, the differences are substantially less pronounced. VFC$_1$ significantly outperforms VFC$_2$ with $p=0.015$. However, neither the comparisons of VFC$_1$ and VFC$_3$ nor VFC$_2$ and VFC$_3$ reaches statistical significance. These results indicate that the choice among automated VFC labeling strategies has limited influence on explanation quality, although manually labeled VFCs provide the strongest overall performance.

% \begin{summarybox}[title=\textbf{Answer to RQ3-1}]
% Manual VFC labels provide the strongest line-level explanation performance, significantly outperforming automated VFC labeling strategies. However, the differences among automated strategies are generally limited, suggesting that further improving automated VFC labeling alone may provide only modest benefits for explanation quality.
% \end{summarybox}

\paragraph{\textbf{VIC labeling strategies}}
The results for VIC labeling, shown in Table~\ref{tab:rq1_vic_mrr}, exhibit a different pattern. VIC$_3$, the LLM-assisted strategy, achieves statistically significant improvement over both VIC$_1$ ($p=0.004$) and VIC$_2$ ($p=0.001$). The comparison between VIC$_3$ and VIC$_0$ does not reach statistical significance ($p=0.102$), but VIC$_3$ has a higher mean MRR.

Overall, LLM-assisted VIC labels significantly improve line-level explanation quality over SZZ-based labels, while also achieving higher MRR than manual labels. These results suggest that LLM-assisted VIC labeling is a promising direction for improving vulnerability explanation and localization.

% Add this to your preamble if not already present:
% \usepackage{threeparttable}
% \usepackage{multirow}
% \usepackage{booktabs}
\begin{table*}[t]
\centering
\caption{Pairwise Wilcoxon signed-rank test results with Cliff's $\delta$ effect sizes of recommended combination, data$_{1-3}$ (VFC$_1$, VIC$_3$) against all other dataset configurations data$_{x-y}$ in the localization task using MRR metric. The bold results show the highest mean performance configurations and statistically significant wins. The $*$ symbol denotes results available only for the Linux kernel.}
\label{tab:rq1_best_combo_mrr}
\begin{threeparttable}
\begin{tabular}{@{}l r r r l r r@{}}
\toprule
\multirow{2}{*}{\textbf{Baselines}} &
  \multirow{2}{*}{\textbf{Mean}} &
  \multirow{2}{*}{\textbf{\textit{p}-value}} &
  \multicolumn{2}{c}{\textbf{Cliff's $\delta$}} &
  \multirow{2}{*}{\textbf{Win\,\%}} &
  \multirow{2}{*}{\textbf{Tie\,\%}} \\
\cmidrule(lr){4-5}
  & & & {\textbf{Value}} & {\textbf{Magnitude}} & & \\
\midrule
data$^*_{1-3}$ & \textbf{0.3855} & -- & -- & -- & -- & -- \\
data$_{1-3}$   & \textbf{0.3980} & -- & -- & -- & -- & -- \\
\midrule
%--- VFC-0 group ---
\multicolumn{7}{@{}l}{} \\[-6pt]
data$^*_{0-0}$  & 0.3828          & 0.2500 & 0.1111 & Negligible & 66.67 & 0.0 \\
data$^*_{1-0}$  & 0.3783 & 0.1250          & 0.5556 & Large & 100.0 & 0.0 \\
data$^*_{2-0}$  & 0.3755 & 0.1250          & 0.5556 & Large      & 100.0 & 0.0 \\
data$^*_{3-0}$  & 0.3797 & 0.2500          & 0.3333 & Medium & 66.67 & 0.0 \\
%--- VFC-1 group ---
\multicolumn{7}{@{}l}{} \\[-6pt]
data$_{0-1}$  & \textbf{0.3980} & 0.2188 & 0.1667 & Small      & 83.33 & 0.0 \\
data$_{1-1}$  & 0.3797 & \textbf{0.0156} & 0.2222 & Small & 100.0 & 0.0 \\
data$_{2-1}$  & 0.3798 & \textbf{0.0156} & 0.2778 & Small      & 100.0 & 0.0 \\
data$_{3-1}$  & 0.3866 & \textbf{0.0469} & 0.1667 & Small  & 83.33 & 0.0 \\

%--- VFC-2 group ---
\multicolumn{7}{@{}l}{} \\[-6pt]

data$_{0-2}$  & 0.3910          & 0.2188 & 0.0556 & Negligible & 50.0  & 0.0 \\
data$_{1-2}$  & 0.3775 & \textbf{0.0156} & 0.2222 & Small & 100.0 & 0.0 \\
data$_{2-2}$  & 0.3823 & \textbf{0.0156} & 0.1667 & Small      & 100.0 & 0.0 \\
data$_{3-2}$  & 0.3838 & \textbf{0.0156} & 0.1667 & Small  & 100.0 & 0.0 \\

%--- VFC-3 group ---
\multicolumn{7}{@{}l}{} \\[-6pt]
data$_{0-3}$  & 0.3957          & 0.2813 & 0.0556 & Negligible & 66.67 & 0.0 \\
data$_{2-3}$  & 0.3888 & \textbf{0.0313} & 0.1111 & Negligible & 83.33 & 0.0 \\
data$_{3-3}$  & 0.3893 & 0.2188          & 0.1667 & Small  & 66.67 & 0.0 \\
\bottomrule
\end{tabular}
\begin{tablenotes}
\footnotesize
\item data$^*_{1-3}$ reports the Linux-only mean for comparisons with Linux-only results \\ data$_{1-3}$ reports the overall mean for other comparisons.
\end{tablenotes}
\end{threeparttable}
\end{table*}

\paragraph{\textbf{Recommended combination.}}
Data$_{1-3}$ achieves the highest observed performance in line-level vulnerability localization, measured by MRR, as shown in Table~\ref{tab:rq1_best_combo_mrr}. It obtains a mean MRR of 0.3980, followed by data$_{0-3}$ and data$_{0-2}$, while tying with data$_{0-1}$. Although data$_{1-3}$ has a comparable mean MRR to data$_{0-1}$, it achieves a higher value in 83.33\% of the paired instances, with no ties. The Linux-only result, data$^*_{1-3}$, is slightly lower at 0.3855, suggesting that the observed performance is not solely driven by a single project. Overall, the MRR results indicate that the localization models can place vulnerable lines relatively early in their ranked explanations, although there remains substantial room for improvement.

Data$_{1-3}$ significantly outperforms seven configurations: data$_{1-1}$, data$_{2-1}$, data$_{3-1}$, data$_{1-2}$, data$_{2-2}$, data$_{3-2}$, and data$_{2-3}$. The largest effect sizes are observed against data$_{1-0}$ and data$_{2-0}$ (Cliff's $\delta=0.5556$), with data$_{1-3}$ winning in 100\% of comparisons. Several other configurations, including data$_{0-1}$, data$_{0-2}$, data$_{0-3}$, and data$_{3-3}$, remain highly competitive, with no statistically significant differences from data$_{1-3}$. Notably, these configurations rely more heavily on accurate labeling, particularly through manual or LLM-assisted labeling. This pattern contrasts with the classification and prioritization results, where balancing label quality and coverage is more important. For localization, label accuracy appears to be more critical, as inaccurate labels can directly mislead the model toward incorrect vulnerable lines.

Overall, data$_{1-3}$ is a promising configuration for vulnerability localization. The differing results across classification, prioritization, and localization further suggest that labeling strategies should be tailored to the downstream task, with localization placing greater emphasis on label accuracy than on label quantity.

\begin{summarybox}[title=\textbf{Answer to RQ3}]
Accurate VFC labels generally improve line-level localization, while differences among automated VFC strategies are limited.
LLM-assisted VIC labeling consistently outperforms SZZ-based labeling and provides competitive localization performance with manual labels, highlighting its potential for improving vulnerability localization.
Overall, data$_{1-3}$ is a promising configuration for vulnerability localization. Unlike classification and prioritization, localization benefits more from label accuracy, suggesting that labeling strategies should be tailored to the downstream task.
\end{summarybox}
% \paragraph{RQ3 summary.}
% Overall, the effects of labeling strategies on explanation quality differ between VFC and VIC labeling. Manual VFC labeling (VFC$_0$) achieves the highest MRR, whereas the LLM-assisted VIC strategy (VIC$_3$) achieves the highest MRR among VIC strategies and significantly outperforms the traditional SZZ-based approaches. Nevertheless, most statistically significant differences have negligible effect sizes, indicating that labeling strategy has a relatively limited practical impact on line-level explanation quality compared with its statistical effect.

\section{Qualitative Analysis}

To complement the quantitative findings, we further conduct a qualitative analysis to investigate the mechanisms underlying the observed task-dependent performance of different labeling configurations. We focus on representative commits where the recommended configurations either perform well or exhibit clear limitations, and examine how the characteristics of the corresponding VFC and VIC labeling strategies shape the supervision available to the models. This analysis helps explain why data$_{0-1}$ is more effective for classification and prioritization, whereas data$_{1-3}$ is more suitable for line-level localization, while also identifying the types of vulnerability patterns for which these advantages weaken.

\label{sec:qualitative}
\subsection{Classification \& Prioritization}
\subsubsection{\textbf{Favourable Cases}}

We focus on these cases because data$_{0-1}$ shows a clear advantage over most alternative configurations on them and because their vulnerabilities arise from subtle interactions in the changed code rather than from explicit security-related cues. Analyzing these examples helps reveal the types of vulnerability patterns that are better preserved by this labeling configuration, and clarifies why its training signal can be more effective for distinguishing and prioritizing vulnerable commits.

In OpenSSL, commit \texttt{3c95ef2}\footnote{https://github.com/openssl/openssl/commit/3c95ef2} introduced Raw Public Key support. The new client path invoked \texttt{ssl\_verify\_rpk()}, but only handled the retry result. A failed verification could therefore be followed by a successful handshake, even when peer authentication was required. The subsequent fix made every unsuccessful verification fatal. Although the inducing commit was presented as a feature extension, the changed code contained the information needed to identify the problem: a verification was performed, but its result was not enforced when updating the connection state.

Linux commit \texttt{3bb3d6b}\footnote{https://github.com/torvalds/linux/commit/3bb3d6b} presents a similar case. It added GMAC support for RK3566 and RK3568 and introduced a shared traversal of
\texttt{ops->regs} until a zero sentinel. This was valid for the newly
populated RK3568 array but not for older platform structures whose flexible array could be empty. The first access could therefore fall outside the valid object. Again, the commit appeared to be ordinary feature development, but the changed code showed both the newly introduced behaviour and the existing data state that made it unsafe.

These cases help explain the advantage of data$_{0-1}$. In both projects, the vulnerability arises from a specific interaction between newly introduced behaviour and an existing program requirement. The OpenSSL case requires enforcing the result of a verification step, whereas the Linux case requires safely handling an existing empty-array state. Such patterns are subtle at the lexical level but are still directly reflected in the changed code.

From the VFC perspective, VFC$_0$ provides a highly curated set of fixes that are closely associated with known vulnerabilities. In contrast, VFC$_2$ prioritizes coverage through broad regular-expression matching, which can include fixes identified from a wider range of security-related messages and therefore with less consistent correspondence between the retrieved commit and the underlying vulnerability mechanism. VFC$_3$ also expands the candidate set through semantic retrieval, but this broader coverage does not necessarily provide a more focused training signal for the specific code interactions observed in these cases. The comparable behaviour of data$_{1-1}$ in the Linux example further suggests that the benefit is related to VFC precision rather than manual annotation alone.

The VIC strategy provides a complementary explanation. VIC$_1$ is vulnerability-specific: it starts from a known vulnerability fix and traces the affected lines through repository history to identify candidate introducing commits. This supervision is well aligned with classification and prioritization, where the model primarily needs to learn whether a commit contains patterns associated with vulnerability introduction rather than identify the exact responsible line. For the favourable cases above, the vulnerable behaviour is closely connected to the code later modified by the fix, allowing VIC$_1$ to preserve a relatively direct relationship between the fixing evidence and the inducing change. This may help explain why data$_{0-1}$ performs reliably on commits whose vulnerability mechanism is visible in the changed code, even when the commit itself appears to be ordinary feature development.

\subsubsection{\textbf{Failure Cases}}

Despite the overall advantage of data$_{0-1}$ in classification and prioritization, a smaller number of commits are better handled by alternative configurations. We examine these cases to characterize the limitations of data$_{0-1}$ and to identify the types of vulnerability patterns for which its advantage diminishes. These exceptions should therefore be interpreted as boundary cases of the recommended configuration, rather than as evidence that the competing configurations perform equally well overall.

OpenSSL commit \texttt{d9ad5b1}\footnote{https://github.com/openssl/openssl/commit/d9ad5b1} attempted to route digests created with \texttt{EVP\_MD\_meth\_new()} through the legacy path by checking
\texttt{type->origin}. The condition appeared reasonable, but initialization could reuse an object already stored in the context when the newly supplied type or cipher was \texttt{NULL}. The subsequent fix therefore had to consider either the new argument or the object retained in the context. All configurations missed this commit during classification, so it is not a classification failure specific to data$_{0-1}$. However, data$_{0-1}$ ranked it lower than several VFC$_2$ based configurations.

Linux commit \texttt{b220c04}\footnote{https://github.com/torvalds/linux/commit/b220c04} shows the same limitation more directly. It added a page size check intended to prevent an overflow in a temporary trace event buffer. However, the calculation reserved space for the trace entry but omitted the additional array element used to store the data length. The commit therefore looked like a genuine safety fix while still allowing an accepted input to exceed the allocated page. In this case, data$_{0-1}$ was weaker in both classification and prioritization.

In both cases, the newly added condition appears locally reasonable, and the vulnerability becomes visible only when additional program context is considered. The OpenSSL case requires tracking an object retained from an earlier call, while the Linux case requires reconstructing the complete buffer layout. These patterns differ from the favourable cases, where the vulnerability mechanism is more directly exposed by the changed statements themselves.

From the VFC perspective, a few alternative configurations, including some using VFC$_2$, perform better on these particular commits. This should be interpreted as a case-specific effect rather than evidence that VFC$_2$ is generally the second-best VFC strategy. One possible explanation is that the broader VFC$_2$ training set exposes the models to a wider range of commits that contain incomplete checks, partial repairs, or security-related changes whose local syntax appears defensive. Such diversity may occasionally help on commits like these, even though the quantitative results show that VFC$_2$ is not consistently competitive overall.

The limitation can also be understood from the VIC perspective. VIC$1$ traces lines affected by a vulnerability fix back through repository history, which provides an effective signal when the inducing behaviour is closely connected to the code later modified by the fix. However, this tracing process does not explicitly reconstruct broader execution state or memory-layout dependencies surrounding the changed lines. Consequently, when the vulnerability depends on information outside the locally modified condition, as in these two cases, the resulting supervision may still identify the relevant inducing commit without fully exposing the contextual relationship that makes it vulnerable. This helps explain why data$_{0-1}$ can lose its advantage on commits whose error is not apparent from the directly changed statements alone.

These failure cases therefore represent a boundary of data$_{0-1}$ rather than a contradiction of its overall advantage. High-quality VFC labels and vulnerability-specific VIC tracing provide a strong commit-level training signal, but they cannot by themselves compensate for model limitations when recognizing vulnerabilities that depend on retained state, implicit object relationships, or complete memory layouts.

\subsection{Localization}
\subsubsection{\textbf{Favourable Cases}}

Linux commit \texttt{3d5048cc}\footnote{https://github.com/torvalds/linux/commit/3d5048cc} moved bridge-link validation into a notifier path. The vulnerable lines do not contain an isolated unsafe operation. Instead, they jointly recover the notifier state, obtain the upper device, and decide whether linking should continue. Under data$_{1-3}$, the attribution is more frequently assigned to the state-extraction lines that connect these steps, whereas data$_{0-1}$ more often favours nearby error-reporting statements. This allows data$_{1-3}$ to rank the lines that implement the vulnerable decision earlier, even though those lines are individually less conspicuous.

OpenSSL commit \texttt{b54cab3}\footnote{https://github.com/openssl/openssl/commit/b54cab3} exhibits a related pattern at a larger scale. It changes the same X509 policy-node operation across several files and call sites. Under data$_{1-3}$, related declarations and calls are more consistently promoted above neighbouring policy-processing statements. The advantage does not come from the size of the change itself, but from the repetition of one semantically coherent operation across the patch.

These cases suggest that data$_{1-3}$ is favourable when the vulnerable lines can be identified through their shared role in a compact decision or a repeated API operation. This behaviour is consistent with the construction of the configuration. VFC$_1$ extends the manually curated base using strict matching criteria, while VIC$_3$ uses semantic information to distinguish commits that actually introduce a vulnerability. Their combination may therefore produce positive examples in which the changed code more consistently reflects the vulnerability-introducing operation. This property is particularly useful for localization, because the model must distinguish the responsible lines from other changed lines within the same commit.

The contrast with data$_{0-1}$ should not be attributed to lower VFC quality. VFC$_0$ is manually curated and achieves the strongest aggregate result among the VFC strategies. The more relevant distinction is the VIC strategy. VIC$_1$ traces lines modified by a fix through repository history and can identify plausible inducing commits without determining whether their changed lines express the actual vulnerability mechanism. Such labels remain effective for learning whether a commit is risky and for prioritizing commits, but they may provide a less direct signal for deciding which line inside a commit is responsible. By incorporating semantic context during VIC identification, VIC$_3$ is better aligned with this fine-grained decision. This interpretation agrees with the quantitative finding that VIC$_3$ improves localization over the SZZ-based alternatives, while differences among automated VFC strategies are comparatively limited.

\subsubsection{\textbf{Failure Cases}}
Linux commit \texttt{20798df}\footnote{https://github.com/torvalds/linux/commit/20798df} shows the boundary of this advantage. Its vulnerable \texttt{d\_inode()} assignment is semantically important but resembles ordinary data preparation. A nearby condition that explicitly checks the response status and dentry state receives greater attribution, causing data$_{1-3}$ to rank the actual vulnerable line later. OpenSSL commit \texttt{e9189cc}\footnote{https://github.com/openssl/openssl/commit/e9189cc} creates a different ambiguity: the vulnerable added condition is almost identical to a deleted condition and mainly differs by the mechanical rename from SSL3\_BUFFER to TLS\_BUFFER. Although data$_{1-3}$ identifies the relevant region, it often prefers the deleted version, while data$_{0-1}$ ranks the vulnerable added line earlier.

These failures indicate that semantic label accuracy does not guarantee exact line ordering. The advantage of data$_{1-3}$ weakens when the responsible line looks generic or is nearly indistinguishable from neighbouring edits. In such cases, visually salient guards or mechanically similar lines can attract more attribution than the true vulnerable line. They also explain why data$_{0-1}$ remains highly competitive: precise manual VFC labels and vulnerability-specific history tracing can still work well when the target line is directly connected to the later fix, and can outperform data$_{1-3}$ when semantic similarity introduces competing candidates.

\section{Discussion}
\label{sec:discussion}
\subsection{Implications}

Based on the findings presented in Section~\ref{sec:evaluation}, we discuss the implications of our study for researchers and outline directions for future research.

% These findings highlight the importance of balancing label quality and coverage when constructing JIT-VP datasets at scale.

\paragraph{\textbf{The choice of labeling techniques is important and task-dependent for dataset construction}}

Our results demonstrate that the choice of labeling strategy is an important design consideration in constructing JIT-VP datasets. Different downstream tasks favor different trade-offs between label quality and coverage, making it unlikely that a single dataset construction strategy will achieve optimal performance across all tasks. For VFC labeling, manual labeling (VFC$_0$) achieves the highest MCC (0.0466) and VDR (0.382), whereas the precision-first strategy (VFC$_1$) achieves classification performance close to the manual baseline (MCC = 0.0434) and a statistically comparable VDR (0.378) for vulnerability prioritization. For VIC labeling, performance differences are generally smaller but remain task-dependent: SZZ-based labeling provides competitive performance for vulnerability prioritization, whereas LLM-assisted labeling (VIC$_3$) is more effective for line-level vulnerability localization. At the dataset configuration level, data$_{0-1}$ achieves the strongest overall performance for classification and vulnerability prioritization, while data$_{1-3}$ performs competitively for line-level vulnerability localization. Although LLM-assisted VIC labeling (VIC$_3$) improves localization performance relative to several traditional SZZ-based configurations, it does not provide the same advantage for vulnerability prioritization. Overall, these findings highlight the varying impacts of labeling techniques and demonstrate that the effectiveness of a labeling strategy depends on the downstream JIT-VP task.

\paragraph{\textbf{Aligning labeling strategies with intended usage provides a path toward practical JIT-VP quality assurance systems.}}

Since the choice of labeling technique is important and task-dependent for dataset construction, our findings suggest that labeling pipelines do not need to follow a one-for-all approach. Instead, different labeling strategies can be combined to build a practical approach for a JIT-VP system. In particular, our results indicate that classification and vulnerability prioritization benefit from a balance between label quality and coverage, whereas line-level vulnerability localization places greater emphasis on label accuracy and semantic detail. Based on these findings, a practical quality-assurance workflow could adopt a two-stage labeling strategy. In the first stage, scalable heuristic or automated techniques could be used to label historical commits, providing both sufficient coverage and reasonable label quality for training classification and vulnerability prioritization models. In the second stage, the initially generated labels could be further refined using more accurate LLM-based or human annotation, particularly for instances requiring precise and semantically detailed labels for vulnerability localization. This sequential labeling process allows a dataset to first achieve a practical balance between quantity and quality and then progressively improve label accuracy where needed. Such a task-specific pipeline could reduce annotation costs and time while maintaining competitive downstream performance, providing a promising direction for developing more efficient and scalable JIT-VP prototypes and, ultimately, practical commit-level quality assurance systems.

\paragraph{\textbf{Capturing the diverse patterns of vulnerable commits remains an open question}}

Our study provides several indications that improving the quality of training labels alone may not be sufficient to substantially improve JIT-VP performance in practical settings. Across our experiments, several changes in labeling strategies result in modest performance differences, particularly for VIC labeling and PR-AUC. Specifically, PR-AUC values range from approximately 0.003 to 0.02 across the evaluated configurations and models, indicating that vulnerability prediction remains challenging under severe class imbalance. Since PR-AUC is sensitive to class prevalence, even a small number of misidentified vulnerable commits in a highly imbalanced dataset can substantially affect precision–recall performance, resulting in very low PR-AUC values. These results suggest that the challenges of realistic JIT-VP extend beyond the quality of training labels. Although future research should continue to improve the quality and coverage of vulnerability labels, equal attention should be given to developing models that can better capture the diverse and subtle patterns associated with vulnerability-introducing changes. In particular, improving models' ability to distinguish vulnerable from non-vulnerable changes under severe class imbalance remains an important open research question.

\subsection{Threats to Validity}

\textbf{Threats to construct validity}. A primary threat to construct validity is the quality of the VFC and VIC labels used to construct the datasets. Both manually curated and automated labeling techniques can introduce false positives and false negatives. In particular, automated approaches may incorrectly associate commits with vulnerabilities or fail to identify relevant vulnerability-related commits. Consequently, differences in model performance across dataset configurations may partly reflect differences in label quality rather than differences in the underlying prediction models. To mitigate this threat, we systematically compare multiple VFC and VIC labeling strategies under the same experimental settings, including the same models, training-data cutoff date, and test sets. We also construct the test sets using the most reliable labels available for each project. For the Linux kernel, we use developer-validated labels to provide higher confidence ground truth. For OpenSSL, where equivalent manually validated labels are unavailable, we rely on vulnerability-related labels derived from established vulnerability references and previously adopted in prior studies~\cite{perl2015vccfinder,nguyen2025toward}. Because this benchmark uses V-SZZ-derived VICs and V-SZZ is also the evaluated implementation of VIC$_1$, the OpenSSL evaluation is not fully independent of that method. We therefore distinguish Linux-only results from analyses that include OpenSSL when interpreting comparisons involving VIC$_1$.

RQ3 additionally requires line-level vulnerability ground truth for evaluating vulnerability localization. We derive vulnerable lines by examining the changes between VIC and VFC and consider a line to be vulnerable when the corresponding line is deleted or modified in the fixing commit. This criterion may not capture vulnerabilities involving interactions among multiple lines, semantic changes, or fixes that substantially restructure the affected code. We therefore retain only commits for which vulnerable lines can be identified with sufficient confidence. The annotation was conducted independently in two stages by two security researchers, each with at least two years of experience with open-source vulnerabilities. The resulting Cohen's Kappa of 0.80 indicates substantial agreement between the annotators.

Although these procedures cannot completely eliminate labeling uncertainty, using consistent test sets across all dataset configurations and explicitly documenting the labeling processes and guidelines helps isolate the effects of the training data labeling strategies on downstream JIT-VP performance and reduce the likelihood that differences in evaluation conditions account for the observed performance differences.

\textbf{Threats to internal validity}. Our conclusions may be affected by the statistical testing procedure and the relatively small number of paired observations available for some comparisons. We use the Wilcoxon signed-rank test because the performance observations cannot necessarily be assumed to follow a normal distribution, and report Cliff's $\delta$ to quantify the magnitude of differences. We consider both statistical significance and effect size rather than relying solely on $p$-values. Nevertheless, the statistical results should be interpreted in conjunction with the observed performance differences.

Particular JIT-VP models and MCR budgets included in the experiments may also affect our conclusions. Although we evaluate multiple ML, DL and LLM models and several labeling configurations, the considered models and budgets represent only a subset of the possible experimental space. Thus, observed differences may vary for other models, hyperparameter settings, or labeling configurations. We reduce this threat by applying the same models, budgets, and evaluation procedures consistently across all dataset configurations.

Another potential threat arises from the difference in the availability of manually curated VIC labels across the two projects. In particular, OpenSSL does not provide the developer-informed references required to construct VIC$_0$. Consequently, configurations using VIC$_0$ are limited to the Linux benchmark, whereas automated VIC strategies can be evaluated on both projects. This difference may introduce project-specific effects into the observed performance and limit the generalizability of our findings. To assess this threat, we use the same manually annotated Linux benchmark across all labeling configurations and compare performance between the Linux-only and cross-project settings as disclosed in the results in Section~\ref{sec:evaluation}. Differences between these results allow us to identify potential project-specific effects and provide evidence for interpreting the robustness of our findings, while also highlighting directions for future evaluation on broader manually curated benchmarks.

\textbf{Threats to external validity}. Our evaluation primarily focuses on the Linux kernel and OpenSSL. Both are mature, security-critical systems-level software with extensive development histories and predominantly C/C++ code. Their development practices, vulnerability characteristics, commit structures, and review processes may differ from those of other software projects. Therefore, our findings may not generalize directly to smaller projects, application software, or projects implemented in other programming languages.

Especially for Linux, we restrict the dataset to commits from 2020 to 2021. This temporal restriction is necessary to construct a manageable and temporally consistent evaluation dataset, but it may limit the representativeness of the identified vulnerability-introducing commits. Based on our validation, the VIC identification process provides approximately 72\% coverage/confidence in capturing the VICs in this period. Consequently, some vulnerability-introducing commits may remain unidentified and are therefore absent from the dataset. Missing VICs can alter the observed vulnerability distribution and potentially affect both model training and evaluation. Thus, the results obtained on Linux should not be interpreted as representative of all vulnerability-introducing commits throughout the entire history of the Linux kernel. We explicitly report this limitation and use the same curated VIC set consistently across experimental configurations to ensure fair comparisons.

\section{Related Work}
\label{sec:related_work}

% In this study, we have reviewed many related works. Here, we provide a summary of the remaining ones.
In this section, we introduce the works in the related fields of JIT-VP, namely: Vulnerability Prediction and Prioritization, Vulnerability Localization and Data/Label Quality in Software Engineering

\paragraph{\textbf{Vulnerability Prediction and Prioritization}}

Effort-aware prediction and ranking is an influential topic in software engineering. Effort-aware defect prediction (EADP) ranks software entities according to their predicted defect likelihood or density so that developers can prioritize inspection under a limited effort budget. Prior studies have extensively investigated learning-to-rank algorithms and demonstrated that the choice of ranking model can substantially affect inspection effectiveness~\cite{yu2019empirical,yu2023finding}. More recent work further argues that treating EADP directly as a ranking problem, rather than as conventional classification or regression, can improve effort-aware performance~\cite{yu2023improving}.

Although these studies demonstrate the value of ranking for software quality assurance, most focus on general defect prediction rather than vulnerability-introducing commits. Moreover, they primarily investigate model and metric design, while the influence of the underlying vulnerability labels used to train ranking models remains less explored. This distinction is particularly important for JIT-VP, where inaccurate or incomplete identification of vulnerability-introducing commits can directly affect the ranking learned by the model and, consequently, the effectiveness of vulnerability prioritization. This sets our study apart from prior work by adopting a data-centric perspective to empirically investigate how different vulnerability labeling strategies affect the quality of downstream JIT-VP models, including their vulnerability prediction and prioritization performance.

\paragraph{\textbf{Vulnerability Localization}}

Recent studies~\cite{pornprasit2022deeplinedp, fu2022linevul} have investigated fine-grained vulnerability localization at the line level to provide developers with actionable information for vulnerability analysis and remediation. For example, learning-based approaches formulate vulnerability localization as a line-level ranking or prediction problem~\cite{hin2022linevd}, while other approaches employ large language models to identify vulnerability-relevant lines~\cite{fu2022linevul, zhou2025large}.

These approaches demonstrate that vulnerability detection and vulnerability localization are related but distinct tasks: a model may successfully identify a vulnerable commit or artifact without accurately identifying the specific lines responsible for the vulnerability. Moreover, a line of code may be syntactically and functionally valid while still containing a security weakness, such as a use-after-free\footnote{https://cwe.mitre.org/data/definitions/416.html} or an Out-of-bounds Write\footnote{https://cwe.mitre.org/data/definitions/787.html} vulnerability. Therefore, accurately identifying vulnerability-introducing lines requires fine-grained labels that capture the specific location and semantic context of the vulnerability, especially for test sets. In contrast to prior studies that primarily focus on developing and evaluating localization models, this study investigates how different labeling strategies affect vulnerability localization performance. In addition, we provide a manually annotated line-level vulnerability dataset and annotation guidelines to facilitate the construction and evaluation of similar datasets in future research.

\paragraph{\textbf{Data and Label Quality in Software Engineering}}

The importance of data quality in software engineering machine learning has been widely recognized~\cite{croft2023data}. Empirical studies have shown that label noise can substantially affect defect prediction models~\cite{tantithamthavorn2015impact,le2024mitigating}. Label quality is particularly important for JIT software defect prediction because labels are often derived indirectly from software repositories. In particular, SZZ and its variants are widely used to identify defect-inducing changes from defect-fixing commits. However, prior studies have shown that SZZ-based labeling can introduce substantial noise and that different SZZ variants can produce different sets of defect-inducing changes~\cite{fan2021impact, herbold2022problems}. This problem also extends to vulnerability prediction. Recent studies have examined automatically labeled vulnerability datasets and found substantial noise in automatically generated labels, while simultaneously showing that noisy labels can still provide useful training signals at scale~\cite{le2024automatic}. These findings illustrate a fundamental trade-off between label quality and labeling scalability.
Recent work has further emphasized that data quality is not simply a preprocessing concern but can influence the conclusions drawn from software engineering models. For example, studies of realistic vulnerability prediction settings have shown that the temporal availability of vulnerability information can introduce additional label noise and substantially affect predictive performance~\cite{garg2022learning}.

Nevertheless, existing research has largely investigated individual sources of label noise or specific labeling algorithms, with limited empirical evidence on how alternative labeling strategies should be selected according to the downstream task. In particular, it remains unclear how different vulnerability labeling strategies affect different JIT-VP tasks, namely vulnerability classification, prioritization, and localization when the same dataset is used for training. This study addresses this gap from a data-centric perspective by systematically evaluating the impact of alternative vulnerability labeling strategies on downstream JIT-VP tasks. By examining classification, vulnerability prioritization, and localization performance together, we provide empirical evidence on the trade-offs between label quality, coverage, and scalability across different downstream usage scenarios.

% \section{Conclusion}
% \label{sec:conclusion}

% We conducted a large-scale empirical study to investigate the impact of data-centric approaches on just-in-time vulnerability prediction (JIT-VP) performance. Our work provides a comprehensive evaluation of multiple labeling configurations across different datasets and model architectures.

% Our findings show that the most effective data configuration is obtained by augmenting manually curated datasets with strictly rule-based VFC labels and LLM-assisted VIC labels. However, the effectiveness of this configuration varies across models, hinting current limitations of existing approaches in capturing the latent characteristics of vulnerability-inducing changes.

% % More broadly, our results highlight the . 
% These findings expose important bottlenecks in JIT-VP research and underscore the need for more robust methods that can effectively leverage imperfect labels while generalizing across projects and model architectures. We hope this work will encourage future research on data-centric methods and contribute to advancing the state of the art in software vulnerability prediction.
% % In the future, we plan to extend our investigations to enable reliable just-in-time prediction for other vulnerability management tasks.

\section{Conclusion}
\label{sec:conclusion}

This paper presents a large-scale empirical study investigating how VFC and VIC labeling strategies affect the performance of JIT-VP models across three downstream tasks: classification, prioritization and localization. We systematically evaluate four VFC and four VIC labeling strategies, resulting in 28 combinations of labeled training data, across two large-scale open-source projects and six representative JIT-VP models. By using a common curated test set across configurations, we provide a consistent comparison under matched experimental conditions of the downstream effects of different data labeling choices.

Our results demonstrate that the choice of labeling strategy can substantially affect downstream JIT-VP performance, and that its effectiveness depends on the task being addressed. Classification and prioritization benefit more from labels that balance quality and quantity, whereas the localization task benefits more from accurate labels. For classification and prioritization, data$_{0-1}$, which combines manual VFC labeling with vulnerability-specific VIC labeling, provides the most reliable overall performance, while precision-first VFC labeling (VFC$_1$) provides competitive performance compared to the recommended option. In contrast, data$_{1-3}$ is among the best-performing configurations for line-level localization. These findings demonstrate that labeling strategies involve different trade-offs between accuracy, coverage, scalability, and cost, and that these trade-offs directly affect the downstream task performance.

More broadly, our findings suggest that, given the available labeling techniques, JIT-VP dataset construction should be viewed as a task-oriented design problem. The same labeling strategy that is effective for identifying and prioritizing vulnerable commits currently does not provide the supervision required for fine-grained localization. This opens a promising direction for constructing task-specific datasets: scalable heuristic or automated labeling can be used to obtain broad coverage for classification and prioritization, while more accurate LLM-assisted or human annotation can be selectively applied when fine-grained supervision is required for localization. Such a strategy can make better use of available annotation effort while retaining competitive downstream performance. At the same time, the limited gains from several labeling improvements indicate that label quality alone cannot fully address the challenges of realistic JIT-VP. Future research should therefore jointly investigate task-specific data construction and models capable of learning diverse vulnerability-specific patterns under severe class imbalance.

% \begin{acks}
% Thank you very much.
% \end{acks}

\bibliographystyle{ACM-Reference-Format}
\bibliography{acm-main}

% \received{6 June 2025}

\end{document}